\documentclass[11pt]{article}

\usepackage{a4wide}
\usepackage{bm}
\usepackage{amsmath,amssymb}
\usepackage{mathrsfs}
\usepackage{comment}
\usepackage{graphicx}
\usepackage{color}
\usepackage{xcolor}
\usepackage{braket}
\usepackage{slashed}
\usepackage{tikz}
\usetikzlibrary{decorations.pathmorphing,decorations.markings,arrows.meta}
\tikzset{
  fermion/.style={postaction={decorate},decoration={markings,mark=at position 0.55 with {\arrow{Stealth[length=5pt]}}}},
  vierbein/.style={double,double distance=1.4pt,decorate,decoration={snake,amplitude=1.5pt,segment length=6pt,pre length=1pt,post length=1pt}},
  mom/.style={-{Stealth[length=4pt]},thin},
}
\usepackage[bookmarks=true,bookmarksnumbered=true,setpagesize=false]{hyperref}
\hypersetup{
	pdftitle={Emergent Einstein-Cartan gravity from a spinor loop},
	pdfauthor={Yadikaer Maitiniyazi, Shinya Matsuzaki, Kin-ya Oda, Yoshiki Uchida, Masatoshi Yamada},
}
\usepackage{enumerate}
\usepackage{pict2e}
\usepackage{booktabs}
\usepackage{mathtools}

\usepackage[nameinlink]{cleveref}

\crefname{section}{Sec.}{Secs.}
\crefname{subsection}{Sec.}{Secs.}
\crefname{appendix}{Appendix}{Appendices}
\crefname{equation}{Eq.}{Eqs.}
\crefname{figure}{Fig.}{Figs.}
\crefname{table}{Tab.}{Tabs.}

\Crefname{section}{Sec.}{Secs.}
\Crefname{subsection}{Sec.}{Secs.}
\Crefname{appendix}{Appendix}{Appendices}
\let\origappendix\appendix
\renewcommand{\appendix}{\origappendix
  \crefalias{section}{appendix}%
  \crefalias{subsection}{appendix}%
}
\Crefname{equation}{Equation}{Equations}
\Crefname{figure}{Fig.}{Figs.}
\Crefname{table}{Tab.}{Tabs.}

\DeclareMathOperator{\tr}{tr}

\DeclareMathOperator{\diag}{diag}

\makeatletter
\DeclareRobustCommand{\intprod}{%
  \mathbin{\mathpalette\int@prod{(0.1,0)(0.85,0)(0.85,0.7)}}%
}
\DeclareRobustCommand{\intprodr}{%
  \mathbin{\mathpalette\int@prod{(0.1,0.7)(0.1,0)(0.85,0)}}}

\newcommand{\int@prod}[2]{%
  \begingroup
  \sbox\z@{$\m@th#1+$}%
  \setlength\unitlength{\wd\z@}%
  \begin{picture}(1,1)
  \roundcap
  \polyline#2
  \end{picture}%
  \endgroup
}
\makeatother

\DeclareMathOperator{\ephogehoge}{\epsilon}
\newcommand{\ep}[1]{\sideset{}{\scriptstyle{[#1]}}\ephogehoge}

\newcommand{\al}[1]{\begin{align}#1\end{align}}

\newcommand{\ov}{\over}
\newcommand{\nn}{\nonumber\\}
\newcommand{\tx}{\text}

\newcommand{\commutator}[2]{\left[#1,\,#2\right]}
\newcommand{\anticommutator}[2]{\left\{#1,\,#2\right\}}
\newcommand{\paren}[1]{\left(#1\right)}
\newcommand{\pn}[1]{\left(#1\right)}
\newcommand{\sqbr}[1]{\left[#1\right]}
\newcommand{\ab}[1]{\left|#1\right|}
\newcommand{\abb}[1]{\left\|#1\right\|}	%

\newcommand{\autospace}{%
  \mathchoice%
    {\!}%
    {\!}%
    {}%
    {}%
}
\newcommand{\fn}[1]{\autospace\paren{#1}} %
\newcommand{\fnl}[1]{\autospace\sqbr{#1}} %
\newcommand{\pa}[1]{\left(#1\right)\!{}}
\newcommand{\mt}[1]{\left[#1\right]\!{}}	%
\newcommand{\Paren}[1]{\bigl(#1\bigr)}
\newcommand{\Pn}[1]{\bigl(#1\bigr)}
\newcommand{\Sqbr}[1]{\bigl[#1\bigr]}

\newcommand{\Fn}[1]{\autospace\Pn{#1}} %

\newcommand{\bs}{\boldsymbol}

\newcommand{\df}{\text{d}}

\newcommand{\I}{I}

\newcommand{\mc}{\mathcal}

\newcommand{\bmat}[1]{\begin{bmatrix}#1\end{bmatrix}}
\newcommand{\pmat}[1]{\begin{pmatrix}#1\end{pmatrix}}

\newcommand{\p}{\partial}

\newcommand{\h}{\hat}
\newcommand{\sth}{\hat\varsigma}%

\newcommand{\ol}{\overline}
\newcommand{\pr}{\prime}

\newcommand{\ba}{\textbf{a}}
\newcommand{\bb}{\textbf{b}}
\newcommand{\bc}{\textbf{c}}
\newcommand{\bd}{\textbf{d}}

\newcommand{\sr}[2]{\stackrel{#1}{#2}}

\newcommand{\os}[2]{\overset{#1}{#2}{}}
\newcommand{\ord}[2]{{\mathstrut}^{(#1)}\!#2}	%

\renewcommand{\t}{\text{t}}

\definecolor{darkgreen}{rgb}{0,0.75,0}
\definecolor{darkred}{rgb}{0.75,0,0}
\definecolor{darkyellow}{rgb}{0.75,0.75,0}
\definecolor{darkcyan}{rgb}{0,0.75,0.75}
\definecolor{darkmagenta}{rgb}{0.75,0,0.75}

\newbox{\ORCIDicon}
\sbox{\ORCIDicon}{\large\includegraphics[width=0.5em]{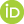}}

\begin{document}
\title{
Emergent Einstein--Cartan gravity from a spinor loop
}\maketitle
\begin{center}

\author{
Yadikaer Maitiniyazi,$^{*\href{https://orcid.org/0009-0004-0826-1130}{\usebox{\ORCIDicon}}}$\footnote{E-mail: \tt ydqem22@mails.jlu.edu.cn}
\
Shinya Matsuzaki,$^{*\href{https://orcid.org/0000-0003-4531-0363}{\usebox{\ORCIDicon}}}$\footnote{E-mail: \tt synya@jlu.edu.cn}
\
Kin-ya Oda,$^{\dagger\href{https://orcid.org/0000-0003-3021-1661}{\usebox{\ORCIDicon}}}$\footnote{E-mail: \tt odakin@lab.twcu.ac.jp}
\\
Yoshiki Uchida,$^{\ddag\href{https://orcid.org/0000-0003-4540-7595}{\usebox{\ORCIDicon}}}$
\footnote{E-mail: \tt uchida.yoshiki@ccnu.edu.cn}
\ and
Masatoshi Yamada$^{\P\href{https://orcid.org/0000-0002-1013-8631}{\usebox{\ORCIDicon}}}$\footnote{E-mail: \tt m.yamada@kwansei.ac.jp}
}
\end{center}
\begin{center}\small\it
$^*$Center for Theoretical Physics and College of Physics, Jilin University, Changchun 130012, China\smallskip\\
$^{\dagger}$Department of Information and Mathematical Sciences, Tokyo Woman’s Christian University, Tokyo 167-8585, Japan\smallskip\\
$^{\ddag}$Institute of Particle Physics and Key Laboratory of Quark and Lepton Physics (MOE), Central China Normal University, Wuhan, Hubei 430079, China\smallskip\\
$^{\ddag}$State Key Laboratory of Nuclear Physics and Technology, Institute of Quantum Matter, South China Normal University, Guangzhou 510006, China\smallskip\\
$^{\ddag}$Guangdong Basic Research Center of Excellence for Structure and Fundamental Interactions of Matter, Guangdong Provincial Key Laboratory of Nuclear Science, Guangzhou 510006, China\smallskip\\
$\P$Department of Physics and Astronomy, Kwansei Gakuin University,
Sanda, Hyogo, 669-1330, Japan
\end{center}
\smallskip

\begin{abstract}\noindent
In Einstein--Cartan spinor gravity under the irreversible vierbein postulate, only the spinor among the first-order variables has a kinetic term of its own at tree level.
Any nonzero background vierbein breaks the local-Lorentz (LL) symmetry spontaneously.
Our previous work showed that the spinor loop generates the kinetic and mass terms of the LL gauge field, manifesting LL as a hidden local symmetry.
Here we show that the spinor loop also generates the vierbein two-point function, while respecting the background gauge invariance and satisfying the Ward--Takahashi identities of the true gauge invariance, for both the general-coordinate and LL symmetries.
The scheme-independent logarithmic divergence assembles the familiar induced-gravity ingredients: a cosmological constant, an Einstein--Hilbert and a Weyl-squared term for the metric sector, and a mass and a higher-derivative kinetic term for the totally antisymmetric torsion.
In the vierbein sector, this torsion is carried by the derivative of the antisymmetric part of the linear vierbein fluctuation.
This part is the would-be Nambu--Goldstone boson of the spontaneously broken LL symmetry, and its couplings in the minimally coupled action are all fixed by general-coordinate and LL gauge invariance.
The five induced terms form a single effective action, manifestly invariant under both symmetries, completing the vierbein sector of the hidden-local-Lorentz emergent-gravity scenario.
\end{abstract}

\newpage
\tableofcontents
\newpage
\section{Introduction %
}
\label{sec:intro}
General relativity describes gravitation at macroscopic scales with extraordinary success, yet after a century its status as a fundamental quantum theory remains unsettled.
This has motivated the search for microscopic descriptions in which gravity is emergent. The graviton may acquire its dynamics from matter loops~\cite{Sakharov:1967pk,Visser:2002ew}, or, more radically, spacetime geometry itself may arise from pregeometric degrees of freedom such as spinors~\cite{Hebecker:2003iw,Wetterich:2003wr}.

Several routes to such an emergent metric description have been explored.
One possibility, going back to the 1970s~\cite{Terazawa:1977xa,Akama:1978collective,Akama:1978unified,Akama:1978pg,Kawati:1979md}, is that the metric tensor is induced by matter degrees of freedom.
In such composite models the graviton is not an elementary particle but a bound state or a collective excitation of more fundamental constituents~\cite{Terazawa:1980vf,Akama:1981dk,Maitiniyazi:2025pvv}.
These scenarios, however, face conceptual challenges. A massless composite graviton is severely constrained by the Weinberg--Witten theorem~\cite{Weinberg:1980kq}, and the microscopic origin and quantum fate of the gauge symmetries of gravity remain long-standing open issues~\cite{Akama:1991np,Akama:1994ehj,Floreanini:1995ie,Chkareuli:2017ufz}.  

Another possibility is that 
the metric description emerges only at long distances, tied to symmetry-breaking in an underlying microscopic theory built on the vierbein as the more fundamental gravitational variable: the pregeometry or spacetime-generation scenario~\cite{Floreanini:1990cf,Percacci:1990wy,Floreanini:1995ie}.
A concrete realization of this idea is provided by the \emph{irreversible vierbein postulate}, based on Einstein--Cartan spinor gravity~\cite{Matsuzaki:2020qzf,Maitiniyazi:2023hts}.
In this scenario, previous studies have mainly focused on the generation of a background spacetime via
the one-loop effective potential induced by spinor fields.
Further work has achieved the simultaneous emergence of the Planck scale
and spacetime geometry~\cite{Maitiniyazi:2024zty}.

Both elements of this scenario have been explored more broadly.
Pregeometry has been realized in various guises, from the early scalar~\cite{Akama:1979tm,Akama:1981kq}, topological~\cite{Akama:1990ur,Akama:1991np,Akama:1994ehj}, and Chern--Simons~\cite{Akama:1991pv} models to recent formulations in which gravity emerges from a symmetry-breaking phase transition~\cite{Volovik:2021wut,Volovik:2023pcm} or from a Yang--Mills-type gauge theory~\cite{Wetterich:2021ywr,Wetterich:2021hru,Wetterich:2021cyp}.
The generation of the Planck scale has likewise been pursued through dynamically broken scale invariance~\cite{Salvio:2014soa,Kubo:2018kho}, no-scale Brans--Dicke gravity~\cite{Hong:2025tyi}, restricted Weyl invariance and $R^2$ gravity~\cite{Edery:2014nha,Edery:2015wha,Edery:2019txq}, spontaneously broken Weyl gauge symmetry~\cite{Ghilencea:2018thl,Ghilencea:2018dqd}, and Weyl-invariant Einstein--Cartan gravity~\cite{Karananas:2024xja}; gravity may even act in the opposite direction and trigger electroweak symmetry breaking~\cite{Shtanov:2023lci}.

Under the irreversible vierbein postulate, only the spinor among the first-order variables has a kinetic term of its own.
Bosonic matter fields, the vierbein, and the local Lorentz (LL) gauge field acquire kinetic terms of their own only through quantum effects generated by spinor fluctuations below an ultraviolet (UV) scale $\Lambda_\tx{G}$.
In Ref.~\cite{Matsuzaki:2020qzf}, this generation has been shown explicitly for the LL gauge field, whereas that for the vierbein has not yet been completed.

In this paper, we extend the previous work~\cite{Matsuzaki:2020qzf} to the dynamical generation of the vierbein kinetic term in Einstein--Cartan spinor gravity under the irreversible vierbein postulate.
In the previous work~\cite{Matsuzaki:2020qzf}, it was shown that once the spacetime background is chosen, the spinor one-loop correction generates the kinetic and mass terms of the LL gauge field, which reveals the LL symmetry as a dynamically generated, spontaneously broken gauge symmetry, i.e., a hidden local symmetry~\cite{Bando:1984ej,Bando:1987br}.
In the present work, it is further clarified that the spinor loop also generates the vierbein two-point function, which respects the background gauge invariance and satisfies the Ward--Takahashi (WT) identities of the true gauge invariance, for both the general-coordinate (GC) and LL symmetries.
Both identities are derived in \cref{app:wti}.
The GC WT identity is confirmed in full by the displayed pole terms.
The LL WT identity is confirmed at zero momentum by the same terms and, at generic momentum, for the $1/\epsilon$ pole with the mixed $e\omega$ amplitude, as described in \cref{sec:ward-takahashi}.

We compute the scheme-independent logarithmic divergence, which assembles the familiar induced-gravity ingredients: It corrects the bare cosmological and Einstein--Hilbert terms, and it adds a Weyl-squared term for the metric sector and a mass and a higher-derivative kinetic term for the totally antisymmetric torsion.
The torsion mass term adds a torsion-squared coupling independent of the Planck mass.
The antisymmetric part of the linear vierbein fluctuation is identified as the would-be Nambu--Goldstone (NG) boson of the spontaneously broken LL symmetry, and its couplings in the minimally coupled action are all fixed by GC and LL gauge invariance.
A polar decomposition of the vierbein removes this mode to all orders.
This completes the vierbein sector of the hidden-local-Lorentz emergent-gravity scenario.
Here ``emergent'' refers to this generation of the bosonic kinetic terms by the spinor loop; the bare Einstein--Cartan term, the term linear in the LL field strength, already yields Einstein--Hilbert dynamics on the nondegenerate branch, and the loop corrects its coefficient.
Our result provides a further step toward understanding how an effective metric description of gravity emerges from an underlying Einstein--Cartan spinor gravity.

This paper is organized as follows. 
In \Cref{Einstein-Cartan spinor gravity}, we provide a brief overview of the construction of Einstein--Cartan spinor gravity. We present the transformation rules of the relevant field components under the LL gauge transformation and the Lie-derivative (LD) form of the GC transformation, and introduce the degenerate limit, the key ingredient of the irreversible vierbein postulate,  
under which the vierbein has no kinetic term of its own at the tree level.
\Cref{Emergence of vierbein kinetic term through loop corrections} 
discusses the emergence of a kinetic term for the vierbein through spinor one-loop corrections, deriving along the way the explicit Feynman rules for the fermion propagator and the interaction vertices.
As a consistency check of the background and true gauge invariance, we derive the WT identities of the true GC and LL symmetries and confirm them for the one-loop result.
In \Cref{sec:induced-action}, we translate the induced one-particle-irreducible (1PI) two-point function into a local effective action, determine the five coefficients, and discuss the implication for the LL-gauge-field sector, in comparison with the literature.
We summarize and discuss our results in \cref{sec:summary}.
\Cref{app:review} collects the derivations and demonstrations supporting \cref{Einstein-Cartan spinor gravity}, \cref{app:frg} describes the functional-renormalization-group (FRG) setting for the propagator-pole question of \cref{sec:two-point}, \cref{app:channels} resolves the 1PI two-point function into spin--parity channels, \cref{app:wti} derives the WT identities of the true GC and LL invariance, \cref{app:polar} establishes the polar decomposition of the vierbein behind the all-orders removal of the would-be NG boson, and \cref{app:heatkernel} reproduces the coefficients of the induced action from the heat-kernel expansion.
\Cref{Einstein-Cartan spinor gravity} and \cref{app:review} follow Refs.~\cite{Matsuzaki:2020qzf,Maitiniyazi:2023hts}.
The new results are presented in \cref{Emergence of vierbein kinetic term through loop corrections,sec:induced-action} and in the appendices that support them, \cref{app:channels-divergences,app:wti,app:polar,app:heatkernel}.

\section{Einstein--Cartan spinor gravity}
\label{Einstein-Cartan spinor gravity}

Our model is based on Einstein--Cartan gravity coupled to spinor fields.
Throughout this work, we refer to this theory as ``Einstein--Cartan spinor gravity.''
Its formulation is presented in detail in Ref.~\cite{Maitiniyazi:2023hts}.
In this section, we summarize the ingredients relevant for the present work, with the remaining review material collected in \cref{app:review}; the reader familiar with Refs.~\cite{Matsuzaki:2020qzf,Maitiniyazi:2023hts} may proceed directly to \cref{Emergence of vierbein kinetic term through loop corrections}.

\subsection{Symmetries}

We start by discussing the symmetries underlying Einstein--Cartan spinor gravity.
We assume that the action at $\Lambda_\tx{G}$ is invariant under the LL gauge transformation and under the GC transformation; the latter is used throughout the main text in its fixed-chart LD form introduced in \cref{sec:LD}.

\subsubsection{LL gauge transformation}
We introduce the vierbein $e^\ba{}_\mu\fn{x}$ and spinor field $\psi\fn{x}$, which transform in the fundamental and spinor representations of the LL gauge symmetry, respectively:
\al{
e^\ba{}_\mu\fn{x}&\sr{\tx{LL}}\to \Lambda^\ba{}_\bb\fn{x}e^\bb{}_\mu\fn{x}, \\
\psi\fn{x}&\sr{\tx{LL}}\to S\Fn{\Lambda\fn{x}}\psi\fn{x},
}
where $\Lambda\fn{x}\in SO(3,1)$, whose infinitesimal form is parametrized as $\Lambda^\ba{}_\bb\fn{x}=\delta^\ba_\bb+\theta^\ba{}_\bb\fn{x}$ with antisymmetric $\theta_{\bb\ba}\fn{x}=-\theta_{\ba\bb}\fn{x}$.
The infinitesimal LL transformation for spinor is given by $S\fn{\Lambda\fn{x}}=1+{1\ov2}\sigma^{\ba\bb}\theta_{\ba\bb}\fn{x}$ with the LL generators $\sigma^{\ba\bb}:=\commutator{\gamma^\ba}{\gamma^\bb}/4$, where the gamma matrices obey the Clifford algebra $\anticommutator{\gamma^\ba}{\gamma^\bb}=2\eta^{\ba\bb}\I$ with the tangent-space metric
\al{
\bmat{\eta_{\ba \bb}}_{\ba,\bb=0,\dots,3} = \mathrm{diag}(-1,1,1,1).
\label{eq:tangentspacemetric}
}
The metric is constructed from the vierbeins in an LL-invariant manner:
\al{
g_{\mu\nu}\fn{x} := \eta_{\ba\bb}\,e^\ba{}_\mu\fn{x} e^\bb{}_\nu\fn{x}.
	\label{composite metric}
}

The LL gauge field $\omega_\mu\fn{x}$ transforms inhomogeneously in the adjoint representation:
\al{
\omega_\mu\fn{x}&\sr{\tx{LL}}\to  \Lambda\fn{x}\omega_\mu\fn{x} \Lambda^{-1}\fn{x} -\Paren{\p_\mu \Lambda\fn{x}} \Lambda^{-1}\fn{x}.
\label{eq:LLgaugetrans}
}
Here we used the short-hand notation for the LL gauge field: $\pa{\omega_\mu}^\ba{}_\bb=\omega^\ba{}_{\bb\mu}$, $\pa{ \Lambda\omega_\mu \Lambda^{-1}}^\ba{}_\bb= \Lambda^\ba{}_\bc\omega^\bc{}_{\bd\mu}\pa{\Lambda^{-1}}^\bd{}_\bb$ with $\pa{\Lambda^{-1}}^\bd{}_\bb=\Lambda_\bb{}^\bd$, etc. 
The field strength tensor is defined by
\al{
\os\omega{\mc F}_{\mu\nu}\fn{x}
	&:=	
    \commutator{\os\omega{\mc D}_\mu}{\os\omega{\mc D}_\nu}
=    \p_\mu\omega_\nu\fn{x}-\p_\nu\omega_\mu\fn{x}
		+\commutator{\omega_\mu\fn{x}}{\omega_\nu\fn{x}},
        \label{eq:LLgaugefieldstrength}
}
where $\os\omega{\mc D}_\mu=\p_\mu + \omega_\mu\fn{x}$ is the covariant derivative associated with the LL gauge transformation on the fundamental representation.
Likewise, the torsion tensor is defined by the antisymmetrized covariant derivative of the vierbein:\footnote{Brackets and parentheses on indices denote antisymmetrization and symmetrization with weight $1/2$, \cref{eq:antisym-notation,eq:sym-notation}.}
\al{
\os{e,\omega}{T}{}^\ba{}_{\mu\nu}\fn{x}:=\os\omega{\mathcal D}_{[\mu} e^\ba{}_{\nu]}\fn{x} = \p_{[\mu} e^\ba{}_{\nu]}\fn{x} + \omega^\ba{}_{\bb[\mu}\fn{x} e^\bb{}_{\nu]}\fn{x}.
\label{torsion defined}
}

\subsubsection{LD transformation}
\label{sec:LD}

The GC invariance is implemented, on the fixed coordinate chart on which the effective action is defined, by the LD transformation along an infinitesimal vector field $\xi^\mu\fn{x}$,
\al{
\delta_\tx{LD}\,\Phi\fn{x}
	&:=	-\mc L_\xi\,\Phi\fn{x},
}
where $\Phi\fn{x}$ stands for any of the fields; the GC transformation itself and its relation to the Lie derivative $\mc L_\xi$ are recalled in \cref{app:review-gc}.
The Lie derivatives of the vierbein, the LL gauge field, and the spinor are given by
\al{
\mc L_\xi\,e^\ba{}_\mu\fn{x}
	&=	\xi\fn{x}e^\ba{}_\mu\fn{x}+\p_\mu\xi^\nu\fn{x}e^\ba{}_\nu\fn{x},\\
\mc L_\xi\,\omega^\ba{}_{\bb\mu}\fn{x}
	&=	\xi\fn{x}\omega^\ba{}_{\bb\mu}\fn{x}+\p_\mu\xi^\nu\fn{x}\omega^\ba{}_{\bb\nu}\fn{x},\\
\mc L_\xi\,\psi\fn{x}
	&=	\xi\fn{x}\psi\fn{x},
}
where $\xi\fn{x}:=\xi^\mu\fn{x}\p_\mu$ denotes the differential operator.

The GC WT identity below is derived with the LD transformation rather than the GC one, for the reason explained in \cref{app:review-gc}.

\subsection{Degenerate limit}
\label{sec:degenerate}

The second assumption is the \emph{irreversible vierbein postulate} that the bare action at $\Lambda_\tx{G}$ must be finite in all of the degenerate limits:
\al{
\abb{e\fn{x}}:=\det_{\ba,\mu}e^\ba{}_\mu\fn{x}\to0,
}
in which some of the parameters $\lambda_a\fn{x}$ introduced below go to zero.

To see this more specifically, we parametrize the $\pn{d+1}^2$ components of the vielbein in $D=d+1$ spacetime dimensions as
\al{
\bmat{e^\ba{}_\mu\fn{x}}_{\ba,\mu=0,\dots,d}
	&=	N\fn{x}\diag\Fn{\lambda_0\fn{x},\dots,\lambda_d\fn{x}}M^\t\fn{x},
	\label{eq: diagonalized vierbein}
}
where $\lambda_a\fn{x}>0$ ($a=0,\dots,d$), while $N\fn{x}\in SO_+(d,1)$ rotates the frame index, $N^\t\eta N=\eta$, and $M\fn{x}\in SO(D)$ rotates the spacetime index, $M^\t M=1$.\footnote{
Both groups are forced by the metric~\labelcref{composite metric}: $g_{\mu\nu}$ is real symmetric, so it is diagonalized by an orthogonal $M$, with eigenvalues $\pn{-\lambda_0^2,\lambda_i^2}$, and the remaining factor $N$ is then automatically Lorentz.
}
The decomposition covers the vielbeins with $\abb{e\fn{x}}>0$, the component that contains the flat background; a vielbein with $\abb{e\fn{x}}<0$ is a frame reflection of one with $\abb{e\fn{x}}>0$ and gives the same metric.

The inverse form of vielbein reads\footnote{
Note that, although the frame index $\ba$ comes first in the symbol $e_\ba{}^\mu\fn{x}$, the subscript ordering $\mu,\ba$ indicates that the rows of this matrix are labeled by the spacetime index $\mu$, so that $M$ again acts on the spacetime index.
}
\al{
\bmat{e_\ba{}^\mu\fn{x}}_{\mu,\ba=0,\dots,d}
	&=	M\fn{x}\diag\fn{{1\ov\lambda_0\fn{x}},\dots,{1\ov\lambda_d\fn{x}}}N^{-1}\fn{x}.
}
In the degenerate limit, some of the $\lambda_a\fn{x}$ are taken to vanish, so that $\abb{e\fn{x}}\to 0$, while $e_\ba{}^\mu\fn{x}\to \infty$.
Our assumption is that the action remains finite in this limit.
This restricts the occurrence of inverse vierbeins in the action without forbidding them altogether: in $D=4$, they are admissible only where the determinant identities~\labelcref{eq: determinant e}--\labelcref{eq: determinant eeee} of \cref{app:review-demos} can eliminate them.
After the elimination, such a term contains no inverse vierbein and is manifestly finite in the degenerate limits.\footnote{
By \cref{eq: determinant e}, each inverse vierbein is a polynomial in the vierbein divided by the determinant, so a local term containing inverse vierbeins is a polynomial divided by a power of the determinant.
The determinant is an irreducible polynomial in the vierbein components, so finiteness in all of the degenerate limits forbids any inverse power of the determinant.
The elimination then proceeds by the repeated use of \cref{eq: determinant e,eq: determinant ee,eq: determinant eee,eq: determinant eeee}.
}

The consequences, demonstrated in \cref{app:review-demos}, are summarized as follows.
Among the fundamental first-order variables, the spinor is the only field with a kinetic term of its own.
Separate kinetic terms for scalar and ordinary gauge fields are forbidden, as are kinetic terms of their own for the symmetric and antisymmetric vierbein components, the latter being torsion-squared terms that require inverse vierbeins.
A vierbein-only Einstein--Hilbert operator is forbidden as well.
Thus the irreversible vierbein postulate highly restricts the action at the UV scale $\Lambda_\tx{G}$.

In a gravitational theory, we expand the vierbein and the LL gauge field around background configurations as
\begin{align}
\label{eq:decomposition}
e^\mathbf{a}{}_\mu\fn{x}
&= \bar{e}^\mathbf{a}{}_\mu\fn{x} +  \hat{e}^\ba{}_\mu\fn{x},
&
\omega^\mathbf{a}{}_{\mathbf{b}\mu}\fn{x}
&= \bar{\omega}^\mathbf{a}{}_{\mathbf{b}\mu}\fn{x} + \hat{\omega}^\mathbf{a}{}_{\mathbf{b}\mu}\fn{x} ,
\end{align}
where fields with a bar and a hat denote the background and fluctuation fields, respectively.
The inverse vierbein is then expanded as
\al{
e_\ba{}^\mu\fn{x}
	&=	\bar e_\ba{}^\mu\fn{x}
-\bar e_\ba{}^\nu\fn{x}
\,\hat{e}^\bc{}_\nu\fn{x}\,
\bar e_\bc{}^\mu\fn{x}
		+\bar e_\ba{}^\nu\fn{x}
\,\hat{e}^\bc{}_\nu\fn{x}\,
\bar e_\bc{}^\rho\fn{x}
\,\hat{e}^\bd{}_\rho\fn{x}\,
\bar e_\bd{}^\mu\fn{x}
+\mathcal{O}\fn{\hat{e}^3}.
\label{eq:expansionInversevierbein}
}
Consequently, gravitational operators containing inverse vierbeins generate infinitely many interaction vertices involving the fluctuation field $\hat{e}^\ba{}_\mu\fn{x}$.
Note that this expansion is well defined only around a nondegenerate vierbein background satisfying $\abb{\bar e\fn{x}}\neq0$.

\subsection{Action}
\label{sec:action}
The irreversible vierbein postulate, together with the minimal truncation specified below, restricts the bare action at $\Lambda_\tx{G}$ to
\begin{align}
\label{eq:Suv}
S_{\Lambda_\tx{G}}
&=
\int\df^4 x\abb{e\fn{x}}
\bigg[
-\Lambda_\tx{cc}
+
\frac{m_\mathrm{P}^2}{2} 
e_\mathbf{a}{}^\mu\fn{x} 
e_\mathbf{b}{}^\nu\fn{x} 
\os\omega{\mc F}^\mathbf{ab}{}_{\mu\nu}\fn{x}
\nn
	&\phantom{=\int\df^4 x\abb{e\fn{x}}\bigg[}
	-{e_\mathbf{a}{}^\mu\fn{x}\ov2}
\pn{
\ol\psi\fn{x} 
\gamma^\mathbf{a}
\os\omega{\mathcal D}_\mu\psi\fn{x}
-
\Pn{\os\omega{\mathcal D}_\mu\ol\psi\fn{x}}
\gamma^\mathbf{a}
\psi\fn{x} 
}
-
m\ol\psi\fn{x}\psi\fn{x}
\bigg],
\end{align}
where the LL field strength $\os\omega{\mc F}^\mathbf{ab}{}_{\mu\nu}\fn{x}$ is defined in \cref{eq:LLgaugefieldstrength} and the LL-covariant derivative is $\os\omega{\mathcal D}_\mu\psi\fn{x}=\p_\mu \psi\fn{x} + \frac{1}{2}\omega_\mathbf{ab\mu}\fn{x}\sigma^\mathbf{ab} \psi\fn{x}$.
Here, $\Lambda_\mathrm{cc}$, $m_\mathrm{P}$, and $m$ are the parameters corresponding to the cosmological constant, the Planck mass, and the spinor mass, respectively; they are bare parameters specified at $\Lambda_\tx{G}$, in the same spirit as the bare Standard-Model parameters at the Planck-scale cutoff~\cite{Hamada:2012bp}.

The action~\labelcref{eq:Suv} is not the most general one compatible with the postulate~\cite{Matsuzaki:2020qzf}.
The Euler, Nieh--Yan, Pontryagin, and Immirzi terms are admissible, but none of them supplies a kinetic term of its own for the vierbein or the LL gauge field; see \cref{app:review-topological}~\cite{Matsuzaki:2020qzf}.
The postulate also admits two parity-even non-minimal fermion--torsion couplings that contain no inverse vierbein,
\al{
S_\text{nm}
	&=	\int\df^4x\,\ep{\mu\nu\rho\sigma}\Bigl[
		\eta_1\,\eta_{\ba\bb}\,\os{e,\omega}{T}{}^\ba{}_{\mu\nu}\fn{x}\,e^\bb{}_\rho\fn{x}\,e^\bc{}_\sigma\fn{x}\,\ol\psi\fn{x}\gamma_\bc\gamma_5\psi\fn{x}
	\nn
	&\phantom{=	\int\df^4x\,\ep{\mu\nu\rho\sigma}\Bigl[}
		+\eta_2\,\ep{\ba\bb\bc\bd}\,\os{e,\omega}{T}{}^\ba{}_{\mu\nu}\fn{x}\,e^\bb{}_\rho\fn{x}\,e^\bc{}_\sigma\fn{x}\,\ol\psi\fn{x}\gamma^\bd\psi\fn{x}
	\Bigr],
	\label{eq:nonminimal-torsion}
}
with $\gamma_5$ the chirality matrix; they supply no bosonic kinetic operator and reduce on the nondegenerate branch, up to normalization, to the axial and trace couplings of Ref.~\cite{Shapiro:2001rz}.
Their parity-odd counterparts, obtained by moving $\gamma_5$ from one current to the other, are admissible as well.

For simplicity we keep only the action~\labelcref{eq:Suv}.
This amounts to a local minimal truncation to operators of canonical dimension at most four in the fundamental first-order variables, with the Hermitian minimally coupled spinor kinetic term.
In the fundamental first-order variables, the action~\labelcref{eq:Suv} contains no kinetic term of its own for either the vierbein or the LL gauge field.
The Einstein--Cartan term mixes the two at first derivative order.
Eliminating the auxiliary connection on the nondegenerate branch then yields Einstein--Hilbert dynamics.
The loop result below therefore corrects the Einstein--Hilbert coefficient rather than generating it.

In Ref.~\cite{Matsuzaki:2020qzf}, we have determined the spinor-loop contributions in the LL-gauge-field sector.
Here we compute the spinor-loop contribution to the vierbein 1PI two-point function.
To this end, we use the decomposition~\labelcref{eq:decomposition} of the vierbein and the LL gauge field into the background and fluctuation parts.

In our scenario, the background vierbein $\bar{e}^\mathbf{a}{}_\mu\fn{x}$ is dynamically generated by quantum fluctuations~\cite{Maitiniyazi:2023hts,Maitiniyazi:2024zty}.
In the present work, however, we leave this generation aside and use the flat configuration
\al{
\bar{e}^\mathbf{a}{}_\mu\fn{x}=\delta^\ba_{\mu},
\label{eq:flatbackground}
}
as the reference around which the spinor loop is computed.
For such a flat background, we may choose the background LL gauge field to vanish, $\bar{\omega}^\mathbf{a}{}_{\mathbf{b}\mu}\fn{x}=0$.
The flat reference need not obey the tadpole condition in the present off-shell calculation; it becomes stationary only after the bare $\Lambda_\tx{cc}$ is tuned so that the full vierbein one-point function vanishes, while the nonzero tadpole retained below supplies the inhomogeneous term in the GC WT identity.
In the next section, we compute the vierbein 1PI two-point function around this background~\labelcref{eq:flatbackground}.

\section{Spinor-induced vierbein two-point function}
\label{sec:emergence}
\label{Emergence of vierbein kinetic term through loop corrections}

In this section, we compute the spinor contribution to the vierbein two-point function.
We evaluate the one-loop diagrams with spinors running in the loop and retain their logarithmically divergent part.
\cref{sec:two-point} computes this part of the vierbein 1PI two-point function, and \cref{sec:ward-takahashi} checks the WT identities.
The divergent part becomes the sole input for the induced action in \cref{sec:induced-action}.

\subsection{One-loop logarithmic divergence}
\label{sec:two-point}
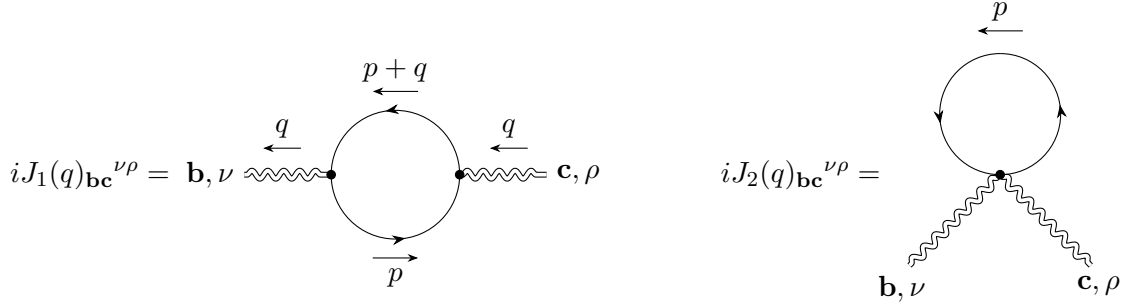
\begin{figure}[t]
\centering
\begin{tikzpicture}
\coordinate (l) at (-0.85,0);
\coordinate (r) at (0.85,0);
\node[left] at (-2.85,0) {$iJ_1(q)_\mathbf{bc}{}^{\nu\rho}=$};
\draw[vierbein] (-2.0,0) -- (l);
\draw[vierbein] (r) -- (2.0,0);
\node[left] at (-2.0,0) {$\bb,\nu$};
\node[right] at (2.0,0) {$\bc,\rho$};
\draw[mom] (-1.25,0.35) -- (-1.75,0.35) node[midway,above]{$q$};
\draw[mom] (1.75,0.35) -- (1.25,0.35) node[midway,above]{$q$};
\draw[fermion] (r) arc (0:180:0.85);
\draw[fermion] (l) arc (180:360:0.85);
\draw[mom] (0.3,1.1) -- (-0.3,1.1) node[midway,above]{$p+q$};
\draw[mom] (-0.3,-1.1) -- (0.3,-1.1) node[midway,below]{$p$};
\filldraw (l) circle (1.6pt);
\filldraw (r) circle (1.6pt);
\coordinate (v) at (8.0,0);
\node[left] at (6.55,0) {$iJ_2(q)_\mathbf{bc}{}^{\nu\rho}=$};
\draw[vierbein] (6.8,-1.2) -- (v);
\draw[vierbein] (v) -- (9.2,-1.2);
\node[below] at (6.7,-1.2) {$\bb,\nu$};
\node[below] at (9.3,-1.2) {$\bc,\rho$};
\draw[fermion] (v) arc (-90:90:0.8);
\draw[fermion] (8.0,1.6) arc (90:270:0.8);
\draw[mom] (8.3,1.9) -- (7.7,1.9) node[midway,above]{$p$};
\filldraw (v) circle (1.6pt);
\end{tikzpicture}
\caption{Spinor one-loop contributions to the vierbein 1PI two-point function. The bubble (left) and seagull (right) diagrams define $iJ_1(q)_{\mathbf{bc}}{}^{\nu\rho}$ and $iJ_2(q)_{\mathbf{bc}}{}^{\nu\rho}$, respectively. The external legs are $\hat e^\mathbf{b}{}_{\nu}(-q)$ and $\hat e^\mathbf{c}{}_{\rho}(q)$, so $q$ enters through the $(\mathbf c,\rho)$ leg and leaves through the $(\mathbf b,\nu)$ leg. Arrows on solid lines indicate fermion flow, and the loop momentum is integrated.\label{fig:2point}}
\end{figure}
At one loop, the spinor contribution to the vierbein 1PI two-point function is the sum of the bubble and seagull diagrams $iJ_1$ and $iJ_2$ of \cref{fig:2point}, and the WT identities of \cref{sec:ward-takahashi} also involve the tadpole $iJ_3$ of \cref{fig:tadpole}, defined in \cref{sec:ward-takahashi}.
The sum of the amputated 1PI diagrams is $i$ times the corresponding variation of the one-loop effective action $\Gamma_\tx{spinor}$, so that
\al{
\Gamma^{(2)ee}|_\tx{spinor}&=J_1+J_2,
\label{eq:Gamma-J12}\\
\Gamma^{(1)e}|_\tx{spinor}&=J_3,
\label{eq:Gamma-J3}
}
where $\Gamma^{(n)\cdots}$ denotes the $n$-th functional derivative of the 1PI effective action with respect to the fluctuations listed in the superscript after the order, here $\hat e$, with the overall momentum-conserving delta function stripped.
The restriction $|_\tx{spinor}$ selects the spinor-loop part, and the unrestricted $\Gamma^{(n)\cdots}$ of \cref{sec:induced-action} and \cref{app:wti} is the full 1PI function.
We work in dimensional regularization with $D=4-2\epsilon$ and keep only the $1/\epsilon$ poles of the one-loop integrals, which carry the scheme-independent logarithmic contribution.
Our Fourier convention is
\al{
f(x)=\int\frac{\df^D p}{(2\pi)^D}\,e^{ip\cdot x}f(p).
}

The relevant interactions are given by
\begin{align}
\label{eq:interactions}
\Delta \mathcal{L}
&=
- 
\abb{\bar{e}\fn{x}}
\bar{e}_{[\mathbf{a}}{}^\mu\fn{x} 
\bar{e}_{{\bf d}]}{}^\sigma\fn{x}
\hat{e}^{\bf d}{}_\sigma\fn{x}
\ol\psi\fn{x}
\gamma^\mathbf{a} 
\overset{\leftrightarrow}{\partial}_\mu 
\psi\fn{x} 
\nonumber\\
&\quad
-\frac{3}{2} 
\abb{\bar{e}\fn{x}}
\bar{e}_{[\mathbf{a}}{}^\mu\fn{x}
\bar{e}_\mathbf{c}{}^\rho\fn{x}
\bar{e}_{\mathbf{d}]}{}^\sigma\fn{x}
\hat{e}^{\bf c}{}_\rho\fn{x}
\hat{e}^{\bf d}{}_\sigma\fn{x}
\ol\psi\fn{x} 
\gamma^\mathbf{a} 
\overset{\leftrightarrow}{\partial}_\mu 
\psi\fn{x} 
\nonumber\\
&\quad
-
\abb{\bar{e}\fn{x}}\bar{e}_\mathbf{d}{}^\sigma\fn{x}
\hat{e}^{\bf d}{}_\sigma\fn{x}
m\ol\psi\fn{x} \psi\fn{x} 
\nonumber\\
&\quad
-
\abb{\bar{e}\fn{x}}\,
\bar{e}_{[\mathbf{c}}{}^\rho\fn{x}
\bar{e}_{\mathbf{d}]}{}^\sigma\fn{x}
\hat{e}^{\bf c}{}_\rho\fn{x}
\hat{e}^{\bf d}{}_\sigma\fn{x}
m\ol\psi\fn{x} \psi\fn{x},
\end{align}
where $A\overset{\leftrightarrow}{\p}_\mu B:=A\p_\mu B-(\p_\mu A)B$.
The Feynman rules follow from \cref{eq:Suv,eq:interactions} as follows.
The spinor propagator carrying momentum $p$ reads
\al{
\begin{tikzpicture}[baseline=-0.6ex]
\draw[fermion] (0,0) -- (2.2,0);
\draw[mom] (0.7,0.3) -- (1.5,0.3) node[midway,above]{$p$};
\end{tikzpicture}
	&=	{1\ov\abb{\bar e}}\,{1\ov-\slashed p+im}.
	\label{eq:spinor-propagator}
}
The vertex with one vierbein fluctuation $\hat e^{\bb}{}_\nu$ carrying an incoming momentum $q$, with an incoming spinor momentum $p_1$ and an outgoing spinor momentum $p_2=q+p_1$, reads
\al{
\vcenter{\hbox{\begin{tikzpicture}[baseline]
\coordinate (v) at (0,0);
\draw[vierbein] (-1.9,0) -- (v);
\node[left] at (-1.9,0) {$\bb,\nu$};
\draw[mom] (-1.6,0.35) -- (-0.9,0.35) node[midway,above]{$q$};
\draw[fermion] (1.2,-1.05) -- (v);
\draw[fermion] (v) -- (1.2,1.05);
\draw[mom] (1.5,-0.75) -- (1.0,-0.31) node[midway,right=2pt]{$p_1$};
\draw[mom] (1.0,0.31) -- (1.5,0.75) node[midway,right=2pt]{$p_2$};
\filldraw (v) circle (1.7pt);
\end{tikzpicture}}}
	&=	{\abb{\bar e}\ov2}
\Sqbr{
\bar e_\bb{}^\nu\pn{\slashed p_1+\slashed p_2-2im}
-\bar e_\bb{}^\mu\gamma^\nu\pn{p_{1\mu}+p_{2\mu}}
}.
	\label{eq:one-vierbein-vertex}
}
The vertex with two vierbein fluctuations $\hat e^{\bb}{}_\nu$ and $\hat e^{\bc}{}_\rho$, with an incoming spinor momentum $p_2$ and an outgoing spinor momentum $p_1$, is independent of the vierbein momenta and reads
\al{
\begin{tikzpicture}[baseline=-0.6ex]
\coordinate (v) at (0,0);
\draw[fermion] (v) -- (-1.0,0.85);
\draw[fermion] (1.0,0.85) -- (v);
\draw[mom] (-0.53,0.09) -- (-0.98,0.47) node[at end,left=1pt]{$p_1$};
\draw[mom] (0.98,0.47) -- (0.53,0.09) node[at start,right=1pt]{$p_2$};
\draw[vierbein] (-1.1,-0.92) -- (v);
\draw[vierbein] (v) -- (1.1,-0.92);
\draw[mom] (-1.06,-0.53) -- (-0.57,-0.11) node[at start,left=1pt]{$q_1$};
\draw[mom] (0.57,-0.11) -- (1.06,-0.53) node[at end,right=1pt]{$q_2$};
\node[below] at (-1.15,-0.92) {$\bb,\nu$};
\node[below] at (1.15,-0.92) {$\bc,\rho$};
\filldraw (v) circle (1.6pt);
\end{tikzpicture}
	&=	\begin{aligned}[t]
		&{\abb{\bar e}\ov2}
\Bigl[
\pn{\slashed p_1+\slashed p_2-2im}
\pn{\bar e_\bb{}^\nu\bar e_\bc{}^\rho-\bar e_\bc{}^\nu\bar e_\bb{}^\rho}
\\&\phantom{{\abb{\bar e}\ov2}\Bigl[}
+\pn{p_{1\mu}+p_{2\mu}}\gamma^\rho
\pn{\bar e_\bb{}^\mu\bar e_\bc{}^\nu-\bar e_\bc{}^\mu\bar e_\bb{}^\nu}
\\&\phantom{{\abb{\bar e}\ov2}\Bigl[}
+\pn{p_{1\mu}+p_{2\mu}}\gamma^\nu
\pn{\bar e_\bc{}^\mu\bar e_\bb{}^\rho-\bar e_\bb{}^\mu\bar e_\bc{}^\rho}
\Bigr].
	\end{aligned}
	\label{eq:two-vierbein-vertex}
}

In the bubble and seagull diagrams of \cref{fig:2point}, $iJ_1(q)_\mathbf{bc}{}^{\nu\rho}$ and $iJ_2(q)_\mathbf{bc}{}^{\nu\rho}$, the momentum $q$ flows into one vierbein leg and out of the other, while the fermion momenta follow the arrows in \cref{eq:one-vierbein-vertex,eq:two-vierbein-vertex}.
In particular, $p_2=p_1+q$ at the one-vierbein vertex.
The amplitudes $iJ_1$ and $iJ_2$ include the standard minus sign from the closed spinor loop.
The displayed one- and two-vierbein vertices are functional derivatives with labeled external legs, so no additional $1/2!$ multiplies either diagram; in particular, the seagull has symmetry factor one.
We evaluate the Dirac traces and the loop integrals analytically and cross-check them with the {\tt FeynCalc} package~\cite{Shtabovenko:2016sxi,Mertig:1990an}.

The logarithmically divergent part of the bubble diagram is
\begin{align}
\label{eq:J1(q)-result}
&iJ_1(q)_{\bf bc}{}^{\nu\rho}\Big|_{1/\epsilon}
\nonumber\\
&=\frac{i}{16\pi^2\epsilon}
\bar{e}_\mathbf{b}{}^\kappa
\bar{e}_{\bf c}{}^\lambda
\bigg[
    \frac{2}{15}
    \left( q^2\eta_{\kappa\lambda} - q_\kappa q_\lambda \right)
    \left( q^2\eta^{\nu\rho} - q^\nu q^\rho \right)
    \nonumber\\
    &\phantom{=\frac{i}{16\pi^2\epsilon}\bar{e}_\mathbf{b}{}^\kappa\bar{e}_{\bf c}{}^\lambda\bigg[}
    -
    \frac{1}{30}
    \left( q^2\delta_\kappa^\nu - q_\kappa q^\nu \right)
    \left( q^2\delta_\lambda^\rho - q_\lambda q^\rho \right)
    -
    \frac{1}{30}
    \left( q^2\delta_\kappa^\rho - q_\kappa q^\rho \right)
    \left( q^2\delta_\lambda^\nu - q_\lambda q^\nu \right)
    \nonumber\\
    &\phantom{=\frac{i}{16\pi^2\epsilon}\bar{e}_\mathbf{b}{}^\kappa\bar{e}_{\bf c}{}^\lambda\bigg[}
    -\frac{1}{3}m^2 
    \Big(
        \pn{q^2\delta_\kappa^\nu - q_\kappa q^\nu} \delta_\lambda^\rho
        +\pn{q^2\delta_\lambda^\nu - q_\lambda q^\nu} \delta_\kappa^\rho\nn
	&\phantom{=\frac{i}{16\pi^2\epsilon}\bar{e}_\mathbf{b}{}^\kappa\bar{e}_{\bf c}{}^\lambda\bigg[-\frac{1}{3}m^2\Big(}
        +\pn{q^2\delta_\kappa^\rho - q_\kappa q^\rho} \delta_\lambda^\nu
        +\pn{q^2\delta_\lambda^\rho - q_\lambda q^\rho} \delta_\kappa^\nu
    \Big)
    \nonumber\\
    &\phantom{=\frac{i}{16\pi^2\epsilon}\bar{e}_\mathbf{b}{}^\kappa\bar{e}_{\bf c}{}^\lambda\bigg[}
    + 
    \frac{2}{3}m^2 
    \pn{q^2\eta^{\nu\rho} - q^\nu q^\rho} \eta_{\kappa\lambda}
    +
    \frac{1}{3}m^2
    	\Bigl(
    	q^2\pn{
			\delta_\kappa^\nu \delta_\lambda^\rho
			+\delta_\kappa^\rho \delta_\lambda^\nu
			}
		-2q_\kappa q_\lambda\eta^{\nu\rho}
		\Bigr)
    \nonumber\\
    &\phantom{=\frac{i}{16\pi^2\epsilon}\bar{e}_\mathbf{b}{}^\kappa\bar{e}_{\bf c}{}^\lambda\bigg[}
    +
    m^4\pn{\delta^\nu_\kappa \delta^\rho_\lambda - \delta_\kappa^\rho \delta_\lambda^\nu}
\bigg] .
\end{align}
The $q^{4}$ and $m^{2}q^{2}$ parts are transverse; only the $m^{4}$ term is not.
The logarithmically divergent part of the seagull diagram is given by
\begin{align}
\label{eq:J2(q)-result}
iJ_2(q)_{\bf bc}{}^{\nu\rho}\Big|_{1/\epsilon}
=
-\frac{i}{8\pi^2\epsilon}
\bar{e}_{\mathbf{b}}{}^\kappa \bar{e}_{\mathbf{c}}{}^\lambda
\pn{\delta_\kappa^\nu \delta_\lambda^\rho - \delta_\lambda^\nu \delta_\kappa^\rho}
m^4 .
\end{align}
Combining \cref{eq:J1(q)-result,eq:J2(q)-result}, it is clear that the vierbein acquires derivative and mass terms through the loop corrections.
In the massless limit of the spinor, the terms proportional to $m^2q^2$ and $m^4$ vanish, and the vierbein 1PI two-point function becomes completely transverse. This behavior is consistent with what is expected from scale invariance.

Here, we take a closer look at the generation of the vierbein kinetic term in the first-order variables of \cref{sec:action}.
In contrast to the claim made in Ref.~\cite{Matsuzaki:2020qzf} that the LL gauge field acquires a kinetic term, the situation for the vierbein is more subtle.
Since the LL gauge field has mass dimension one whereas the vierbein is dimensionless, the two-point function of the vierbein exhibits a $q^4$ dependence that is absent in the LL gauge field case.
Although it is not yet clear what should properly be regarded as the kinetic term for the vierbein, in this paper we shall use the term {\it kinetic term} in a broader sense to refer to any two-point self-interaction involving derivatives.

Before concluding this section, we briefly comment on the difference between the dynamical emergence of the LL gauge field discussed in our previous work~\cite{Matsuzaki:2020qzf} and that of the vierbein.
For the LL gauge field, two types of kinetic terms appear: one for the transverse mode, which remains finite at the one-loop level, and the other for the longitudinal mode, which exhibits a logarithmic divergence.
In contrast, for the vierbein, only the kinetic term for the transverse mode is generated, and it is logarithmically divergent.\footnote{
How this logarithmic divergence is interpreted in terms of the physical cutoff $\Lambda_\tx{G}$ is explained in \cref{sec:result}.
}
This observation suggests that the vierbein is fundamentally different from the LL gauge field in its dynamical behavior.

At one spinor loop, the termination at $q^4$ is exact, not an artifact of truncating an expansion. The $1/\epsilon$ pole of a one-loop diagram is a local polynomial of mass dimension four in $m$ and $q$, so any term beyond $q^4$ would require inverse powers of $m$ and cannot appear in the divergence.
The structures displayed in \cref{eq:J1(q)-result,eq:J2(q)-result} are therefore exhaustive.
The finite part is not constrained in this way. It is a nontrivial function of $q^2/m^2$, so that finite corrections to the kinetic terms arise at all orders in $q^2$.

We have shown that the spinor one-loop contribution generates $q^0$, $q^2$ and $q^4$ terms in the vierbein 1PI two-point function.
At first sight, this may raise the concern that the vierbein propagator contains an Ostrogradsky ghost~\cite{Ostrogradsky:1850fid,Woodard:2015zca}, which in the metric sector is the Stelle ghost of quadratic gravity~\cite{Stelle:1976gc}.\footnote{The classical chart of when quadratic-torsion multiplets are ghost- and tachyon-free is given in Ref.~\cite{Hayashi:1980qp}.}
More specifically, the propagator may schematically take the form
\al{
\frac{1}{q^4+M^2q^2}
=
\frac{1}{M^2}\left(
\frac{1}{q^2}
-
\frac{1}{q^2+M^2}
\right),
}
where the second term has the wrong sign and is therefore interpreted as the propagator of a negative-norm Ostrogradsky mode.

However, the present one-loop calculation is not sufficient to conclude that such a negative-norm pole actually exists. The full quantum effective action generally acquires a nontrivial momentum dependence beyond the polynomial terms obtained in perturbation theory, and the analytic structure of the propagator can therefore be substantially modified. To address this issue, it is useful to examine the problem within the FRG framework described in \cref{app:frg}.

The resolution of \cref{eq:J1(q)-result,eq:J2(q)-result} into spin--parity channels is recorded in \cref{app:channels}.

\subsection{Ward--Takahashi identities}
\label{sec:ward-takahashi}
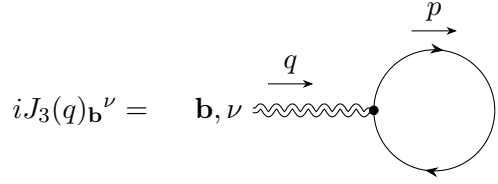
\begin{figure}[t]
\centering
\begin{tikzpicture}
\coordinate (v) at (0,0);
\node[left] at (-2.85,0) {$iJ_3(q)_\mathbf{b}{}^{\nu}=$};
\draw[vierbein] (-1.6,0) -- (v);
\node[left] at (-1.6,0) {$\bb,\nu$};
\draw[mom] (-1.4,0.35) -- (-0.8,0.35) node[midway,above]{$q$};
\draw[fermion] (v) arc (180:0:0.8);
\draw[fermion] (1.6,0) arc (0:-180:0.8);
\draw[mom] (0.5,1.05) -- (1.1,1.05) node[midway,above]{$p$};
\filldraw (v) circle (1.6pt);
\end{tikzpicture}
\caption{Spinor one-loop vierbein tadpole defining $iJ_3(q)_\mathbf{b}{}^\nu$. The external leg $\hat e^\mathbf{b}{}_\nu(q)$ carries momentum $q$ into the fermion loop; the loop momentum is integrated and the overall $(2\pi)^D\delta^{(D)}(q)$ is suppressed. Arrows on the solid line indicate fermion flow.\label{fig:tadpole}}
\end{figure}

The effective action is invariant under the true GC and LL transformations.
Each invariance gives a WT identity for the vierbein two-point function at linear order in $\hat{e}$.
Both identities are derived in \cref{app:wti}.
Both contain the vierbein tadpole, so we calculate it first.
The tadpole diagram is shown in \Cref{fig:tadpole} and denoted by $iJ_3(q)_\mathbf{b}{}^{\nu}$.
The logarithmically divergent part of the tadpole diagram is given by
\begin{align}
\label{eq:J3(q)-result}
iJ_3(q)_\mathbf{b}{}^{\nu}
\Big|_{1/\epsilon}
=
-
\bar{e}_\mathbf{b}{}^\nu
\frac{im^4}{16\pi^2\epsilon} .
\end{align}

The GC WT identity is derived with the LD transformation of \cref{sec:LD}, which implements the GC invariance on the fixed chart.
We confirmed that the GC WT identity holds at the linear order of $\hat{e}$:
\begin{align}
\label{eq:wi_wi-3}
q_\nu^{}  
\bar{e}^\mathbf{b}{}_\sigma
\Big( 
J_1(q)_{\bf bc}{}^{\nu\rho} + J_2(q)_{\bf bc}{}^{\nu\rho} 
\Big)
\Big|_{1/\epsilon}
&=
\Big(
q_\sigma^{}
J_3(q)_\mathbf{c}{}^\rho
-
q_\nu^{}
J_3(q)_\mathbf{c}{}^\nu
\delta_\sigma^\rho
\Big)
\Big|_{1/\epsilon}
;
\end{align}
see \cref{app:wti-ld} for the derivation.
At $q=0$ the two-point function is a linear combination of only three tensors, $\bar{e}_\mathbf{b}{}^\nu\bar{e}_\mathbf{c}{}^\rho$, $\bar{e}_\mathbf{b}{}^\rho\bar{e}_\mathbf{c}{}^\nu$, and $\eta_\mathbf{bc}\eta^{\nu\rho}$.
\Cref{eq:wi_wi-3} then determines the coefficients of all three tensors from the tadpole $J_3$.
At order $q^2$ and $q^4$ the right-hand side of \cref{eq:wi_wi-3} vanishes, so the two-point function is transverse.

The LL WT identity, derived in \cref{app:wti-ll}, reads at one loop
\al{
\label{eq:wi_ll-3}
&\bar{e}_{\mathbf{b}\nu}
\Big(
J_1(q)_{\bf ac}{}^{\nu\rho} + J_2(q)_{\bf ac}{}^{\nu\rho}
\Big)
-
\bar{e}_{\mathbf{a}\nu}
\Big(
J_1(q)_{\bf bc}{}^{\nu\rho} + J_2(q)_{\bf bc}{}^{\nu\rho}
\Big)
+
iq_\nu
J^{e\omega}(q)_{\mathbf{c},\mathbf{ab}}{}^{\rho\nu}
\Big|_{1/\epsilon}
\nn
&=
\Big(
\eta_{\bf ac}
J_3(q)_\mathbf{b}{}^\rho
-
\eta_{\bf bc}
J_3(q)_\mathbf{a}{}^\rho
\Big)
\Big|_{1/\epsilon},
}
where $iJ^{e\omega}$ is the amputated one-loop mixed $e\omega$ two-point function, normalized as in \cref{eq:wti_Gamma-ll}, with the comma separating the vierbein index from the index pair of the LL gauge field.
The identity relates the antisymmetric part of the $ee$ two-point function to the tadpole and to the mixed amplitude.
The mixed term starts at order $q^2$, as shown in \cref{app:wti-ll}.
At $q=0$ it drops out, and both sides equal $\tfrac{m^4}{16\pi^2\epsilon}\pn{\eta_{\bf bc}\bar{e}_\mathbf{a}{}^\rho-\eta_{\bf ac}\bar{e}_\mathbf{b}{}^\rho}$ by \cref{eq:J1(q)-result,eq:J2(q)-result,eq:J3(q)-result}.
This agreement follows from \cref{eq:wi_wi-3}, which fixes the $q=0$ part of the two-point function.
At order $q^2$ and $q^4$ the mixed amplitude enters, and the LL WT identity is no longer a consequence of the GC WT identity.
We computed the mixed amplitude and confirmed \cref{eq:wi_ll-3} for the $1/\epsilon$ pole at generic $q$ in all 96 components, six pairs $\mathbf{ab}$ times sixteen legs $\mathbf{c}\rho$.
This check plays no role in determining the coefficients in \cref{eq:induced-action}.
The closed form of $J^{e\omega}$ and the convention dictionary are deferred to the mixed-sector analysis in Ref.~\cite{Oda:2026prep}.

\section{Low energy effective theory}
\label{sec:induced-action}

We now determine the local induced action that reproduces the 1PI two-point function of \cref{sec:two-point}.
That two-point function fixes five coefficients, up to terms whose second variation vanishes around the flat reference.
The construction is organized by LL invariance:
\cref{sec:trade} trades the elementary pair $(e,\omega)$ for the composite metric and the contorsion;
\cref{sec:result} states the resulting induced action;
\cref{sec:determination} determines its coefficients from the loop, with the antisymmetric vierbein entering as the linearized LL NG mode;
and \cref{sec:discussion} gives the induced-gravity reading, the prediction for the LL-gauge-field sector, and the relation to the literature.

\subsection{The change of variables}
\label{sec:trade}

Below $\Lambda_\tx{G}$ the vierbein is generically invertible, which allows one to write down the composite Levi-Civita spin connection
\al{
\os{e}{\Omega}{}^\ba{}_{\bb\mu}\fn{x}
	&:=	e^\ba{}_\lambda\fn{x}\os g\nabla_\mu e_\bb{}^\lambda\fn{x},
	\label{eq:composite-Omega}
}
where the GC-covariant derivative acts only on the spacetime index,
\al{
\os g\nabla_\mu e_\bb{}^\lambda\fn{x}
	&=	\p_\mu e_\bb{}^\lambda\fn{x}+\os g\Gamma^\lambda{}_{\rho\mu}\fn{x}e_\bb{}^\rho\fn{x},
		\label{GC-covariant derivative}
}
with the Levi-Civita connection of the composite metric
\al{
\os g\Gamma^{\lambda}{}_{\rho\mu}\fn{x}
	:=	{g^{\lambda\gamma}\fn{x}\ov2}
		\Pn{-\p_\gamma g_{\rho\mu}\fn{x}+\p_\rho g_{\mu\gamma}\fn{x}+\p_\mu g_{\gamma\rho}\fn{x}}.
	\label{eq:Levi-Civita}
}
The composite decoration follows the pattern of $\os{\omega}{\mc F}$ in \cref{eq:LLgaugefieldstrength}, and keeps the connection~\labelcref{eq:composite-Omega} distinct from the theory's \emph{elementary} connection $\omega$.

Two properties follow directly from the definition~\labelcref{eq:composite-Omega}.
First, the Levi-Civita spin connection is antisymmetric in its frame pair, $\os{e}{\Omega}_{\ba\bb\mu}\fn{x}=\os{e}{\Omega}_{[\ba\bb]\mu}\fn{x}$, because $\os g\nabla$ preserves the metric,
\al{
\os{e}{\Omega}_{(\ba\bb)\mu}\fn{x}
	=	-\frac12\,e_{\ba}{}^{\nu}\fn{x}\,e_{\bb}{}^{\rho}\fn{x}\,\os g\nabla_{\mu}^{}g_{\nu\rho}\fn{x}
	&=	0.
	\label{metricity}
}
Second, unlike the elementary torsion~\labelcref{torsion defined}, the torsion of the composite connection vanishes\footnote{
\cref{eq:composite-connection} is not a metricity condition.
Its first equality is the antisymmetrized part of the identity $\partial_\mu e^{\ba}{}_\nu+\os{e}{\Omega}^{\ba}{}_{\bb\mu}e^{\bb}{}_\nu=e^{\ba}{}_\lambda\os g\Gamma^{\lambda}{}_{\nu\mu}$, which follows from the definition~\labelcref{eq:composite-Omega} and states that the vierbein is covariantly constant when the Levi-Civita connection acts on its spacetime index and $\os{e}{\Omega}$ on its frame index.
The second equality holds because the Christoffel symbol is symmetric.
The $4\times6=24$ components of \cref{eq:composite-connection} count the torsion, whereas metricity is the $40$-component statement~\labelcref{metricity}.
}
\al{
\p_{[\mu}^{}e^{\ba}{}_{\nu]}\fn{x}
	+\os{e}{\Omega}^{\ba}{}_{\bb[\mu}\fn{x}\,e^{\bb}{}_{\nu]}\fn{x}
	=	e^{\ba}{}_{\lambda}\fn{x}\,\os g\Gamma^{\lambda}{}_{[\nu\mu]}\fn{x}
	&=	0.
	\label{eq:composite-connection}
}
These two properties fix the composite connection uniquely~\cite{Hehl:1976kj,Shapiro:2001rz}.

On the invertible branch $\abb{e\fn{x}}\neq0$ the field content trades exactly,
\al{
\underbrace{16}_{e}+\underbrace{24}_{\omega}
	&=	\underbrace{6}_{\Lambda}
		+\underbrace{10}_{g}
		+\underbrace{24}_{K},
			\label{eq:GammagK}
}
where
\al{
K^{\ba}{}_{\bb\mu}\fn{x}
	&:=	\omega^{\ba}{}_{\bb\mu}\fn{x}-\os{e}{\Omega}{}^{\ba}{}_{\bb\mu}\fn{x}
		\label{contorsion}
}
is the contorsion~\cite{Hehl:1976kj,Shapiro:2001rz,Hayashi:1980ir},\footnote{
For the metric $g$ and the contorsion $K$, and only for these, we refrain from the composite decorations $\os{e}{g}$ and $\os{e,\omega}{K}$, in the spirit of the change of variables~\labelcref{eq:GammagK}.
}
which transforms homogeneously under the LL transformation,
\al{
K^\ba{}_{\bb\mu}\fn{x}
	&\sr{\tx{LL}}\to	\Lambda^\ba{}_\bc\fn{x}\,K^\bc{}_{\bd\mu}\fn{x}\,\pa{\Lambda^{-1}}^\bd{}_\bb\fn{x},
	\label{eq:K-LL}
}
because the composite connection~\labelcref{eq:composite-Omega} transforms inhomogeneously, exactly the same as the elementary $\omega$ in \cref{eq:LLgaugetrans}, and the inhomogeneous shift cancels in the difference~\labelcref{contorsion}.
The 16 components of an invertible vierbein carry the 10 of the composite metric $g$ together with the 6-parameter LL transformation $\Lambda$.
Two vierbeins compose the same metric precisely when they differ by an LL transformation $e^{\prime\,\ba}{}_{\mu}\fn{x}=\Lambda^{\ba}{}_{\bb}\fn{x}\,e^{\bb}{}_{\mu}\fn{x}$:
\al{
\eta_{\ba\bb}\,e^{\prime\,\ba}{}_{\mu}\fn{x}\,e^{\prime\,\bb}{}_{\nu}\fn{x}
	=	\eta_{\ba\bb}\,e^{\ba}{}_{\mu}\fn{x}\,e^{\bb}{}_{\nu}\fn{x}.
	\label{eq:LL-fiber}
}

This split of the 16 vierbein components into the composite metric and the LL orientation admits an explicit all-orders parametrization.
Near the background, the vierbein factors uniquely into an LL element, the background, and a symmetric stretch (the stretch tensor, in the language of continuum mechanics),\footnote{
In Euclidean signature this would be, for the compound matrix $e^\ba{}_\bb:=e^{\ba}{}_{\mu}\bar e_{\bb}{}^{\mu}$, nothing but the ordinary polar decomposition: an orthogonal $SO(4)$ rotation times a symmetric positive-definite factor.
}
\al{
e^{\ba}{}_{\mu}\fn{x}
	&=	\pa{e^{\h\kappa\fn{x}}}^{\ba}{}_{\bb}\,
		\bar e^{\,\bb}{}_{\nu}\fn{x}\,
		\pn{\delta^{\nu}_{\mu}+\frac12\sth{}^{\nu}_{\mu}\fn{x}},
	\label{eq:polar-decomposition}
}
where $\h\kappa_{\ba\bb}\fn{x}=\h\kappa_{[\ba\bb]}\fn{x}$ is the would-be NG boson and $\sth_{\mu\nu}\fn{x}=\sth_{(\mu\nu)}\fn{x}$ is the stretch.\footnote{
Throughout, spontaneous breaking of the local LL symmetry and the associated NG language are perturbative Higgs-phase shorthand in the chosen background expansion.
Elitzur's theorem forbids a gauge-variant local order parameter from acquiring an expectation value in the unfixed formulation~\cite{Elitzur:1975im}; the gauge-invariant content used here is the St\"uckelberg organization of the LL orientation and the invariant metric/contorsion variables, not a literal breaking of the gauge redundancy.
}
Spacetime indices are raised and lowered with the background metric, and frame and spacetime indices are converted with the background vierbein.
The count matches \cref{eq:GammagK}, with the $6$ of $\h\kappa$ carrying the LL orientation and the $10$ of $\sth$ carrying the metric.
Since an LL transformation acts on the Lorentz factor alone, the would-be NG boson can be gauged away entirely, $\h\kappa\fn{x}=0$ to all orders rather than only at linear order; in this unitary gauge its LL orientation is absorbed into the St\"uckelberg combination with the connection.
The uniqueness of the factorization~\labelcref{eq:polar-decomposition}, the expansions of the vierbein fluctuation in the polar fields, and the transformation laws behind these statements are established in \cref{app:polar}.

Of the traded variables~\labelcref{eq:GammagK}, the LL orientation $\Lambda$ drops out, because the fermion measure is LL-anomaly-free in four dimensions~\cite{Alvarez-Gaume:1983ihn,Bardeen:1984pm}.
The induced action is therefore a functional $\Gamma_\tx{ind}\fnl{\bar g,\bar K}$ of the remaining 34 components, which enter through invariant contractions.
The construction presupposes the invertible branch. Neither $\os{e}{\Omega}$ nor $K$ survives the degenerate limit of \cref{sec:degenerate}, and the induced action below is the theory's output on that branch.

\subsection{The spinor-induced action}
\label{sec:result}

We can now state one of the main results of this paper.
The vierbein 1PI two-point function of \cref{sec:two-point} fixes the following spinor-one-loop induced action uniquely, up to operators\footnote{
As usual in effective field theory, ``operator'' means a local monomial in the (background) fields and their derivatives.
} invisible to its $ee$ two-point function around the flat background:
\al{
\Gamma_\tx{ind}\fnl{\bar g,\bar K}
	&=	-{1\ov16\pi^{2}\epsilon}\int\df^{4}x\,
		\abb{\bar e\fn{x}}\bigg[ m^{4} + {m^{2}\ov6}\,\os{\bar g}R\fn{x}
		- {1\ov40}\,\os{\bar g}C^{2}\fn{x}
	\nn
	&\phantom{=	-{1\ov16\pi^{2}\epsilon}\int\df^{4}x\,\abb{\bar e\fn{x}}\bigg[}
		- {3\ov2}\,m^{2}\,\bar K_{[\lambda\mu\nu]}\fn{x}\bar K^{[\lambda\mu\nu]}\fn{x}
		\nn&\phantom{=	-{1\ov16\pi^{2}\epsilon}\int\df^{4}x\,\abb{\bar e\fn{x}}\bigg[} - {1\ov4}\,\os{\bar g}\nabla_{\rho}^{}\bar K_{[\lambda\mu\nu]}\fn{x}\,
		\os{\bar g}\nabla^{\rho}\bar K^{[\lambda\mu\nu]}\fn{x} \bigg],
	\label{eq:induced-action}
}
where
\al{
\os g C^{2}\fn{x}:=\os g C_{\mu\nu\rho\sigma}\fn{x}\os g C^{\mu\nu\rho\sigma}\fn{x} =\os g R_{\mu\nu\rho\sigma}\fn{x}\os g R^{\mu\nu\rho\sigma}\fn{x}-2\os g R_{\mu\nu}\fn{x}\os g R^{\mu\nu}\fn{x}+\frac13\os g R^{2}\fn{x}
}
is the square of the Weyl tensor (the second equality holds in $D=4$) and $\bar K_{[\lambda\mu\nu]}\fn{x}$ the totally antisymmetric part of the contorsion~\labelcref{contorsion}, evaluated on the background. Recall that $m$ is the spinor mass in \cref{eq:Suv}.
A single spinor loop thus yields the five logarithmic contributions in \cref{eq:induced-action} with fixed relative coefficients.

The logarithmic divergence of the one-loop integrals appears in dimensional regularization as the pole $1/\epsilon$ in \cref{eq:induced-action}.
Here the bare action~\labelcref{eq:Suv} is specified at the physical scale $\Lambda_\tx{G}$, and the effective action is evaluated at a lower scale $\mu$.
With the spinor loop restricted to momenta between $\mu$ and $\Lambda_\tx{G}$, the pole becomes the finite logarithm $\ln(\Lambda_\tx{G}^{2}/\mu^{2})$, which is positive for $\mu<\Lambda_\tx{G}$.
The first two of the five contributions then correct the bare cosmological and Einstein--Hilbert coefficients in \cref{eq:Suv}.
The shifts are $+(m^{4}/16\pi^{2})\ln(\Lambda_\tx{G}^{2}/\mu^{2})$ for $\Lambda_\tx{cc}$ and $-(m^{2}/48\pi^{2})\ln(\Lambda_\tx{G}^{2}/\mu^{2})$ for $m_\mathrm{P}^{2}$, so the spinor loop raises the cosmological constant and lowers the Planck mass below $\Lambda_\tx{G}$.

The last three terms correct no coefficient of \cref{eq:Suv}.
The Weyl-squared term and the kinetic term of the totally antisymmetric contorsion $\bar K_{[\lambda\mu\nu]}$, the third and fifth terms of \cref{eq:induced-action}, have no counterpart there at all.
The mass term of $\bar K_{[\lambda\mu\nu]}$, the fourth term, has a partial one: written in the metric and the contorsion, the Einstein--Cartan term contains torsion-squared operators, but only in one fixed combination with the single coefficient $m_\mathrm{P}^{2}$.
The loop generates the totally antisymmetric channel alone, which is not a rescaling of that combination, so the low-energy action acquires a torsion-squared coupling independent of $m_\mathrm{P}^{2}$.

At first glance, these three terms might appear to require counterterms that the postulate forbids.
A counterterm would be needed only if the cutoff were sent to infinity, whereas $\Lambda_\tx{G}$ is a physical scale here and stays finite.
Rather, the postulate sets their coefficients to zero at $\Lambda_\tx{G}$, and the logarithm gives them calculable values at lower scales on a nondegenerate background.
\Cref{eq:induced-action} determines only this logarithmic part.
The power-sensitive and finite parts, and with them the full low-energy couplings, are left undetermined here and are discussed in \cref{sec:summary}.

Only four of the 24 components of the contorsion~\labelcref{contorsion}, its totally antisymmetric part $K_{[\lambda\mu\nu]}$, enter the two contorsion terms of \cref{eq:induced-action}; the other 20, the trace vector $K^{\lambda}{}_{\mu\lambda}$ and the mixed-symmetry part, are absent from the minimally coupled spinor action.\footnote{\label{contorsion decomposition}
Explicitly, $K_{\lambda\mu\nu}=K_{[\lambda\mu\nu]}+\tfrac13\bigl(g_{\lambda\nu}K^{\rho}{}_{\mu\rho}-g_{\mu\nu}K^{\rho}{}_{\lambda\rho}\bigr)+q_{\lambda\mu\nu}$, where the mixed-symmetry part $q_{\lambda\mu\nu}$ is antisymmetric in $\lambda\mu$, traceless, and has no totally antisymmetric part; the three pieces carry $4+4+16$ components.
}
In the symmetrized kinetic term~\labelcref{eq: kinetic term of spinor}, the LL gauge field~$\omega$ appears only through
\al{
e_{\ba}{}^{\mu}\fn{x}\,\omega_{\bb\bc\mu}\fn{x}
\anticommutator{\gamma^{\ba}}{\sigma^{\bb\bc}}.
}
The anticommutator is totally antisymmetric, $\anticommutator{\gamma^{\ba}}{\sigma^{\bb\bc}}=\gamma^{[\ba}\gamma^{\bb}\gamma^{\bc]}$, so only $\omega_{[\ba\bb\bc]}\fn{x}$ enters the spinor loop~\cite{Hehl:1976kj,Shapiro:2001rz}.
Operators beginning beyond quadratic order about the flat, torsionless reference, such as $\os{\bar g}R\bar K\bar K$ and $\bar K^4$, are invisible to the present two-point function.
The parity-odd Nieh--Yan term~\labelcref{Nieh-Yan action} is likewise invisible up to a total derivative.

All five coefficients follow from the vierbein 1PI two-point function alone through the matching of \cref{sec:determination}, and an independent covariant heat-kernel calculation with the standard Schwinger--DeWitt coefficients~\cite{Vassilevich:2003xt} reproduces all five, as shown in \cref{app:heatkernel}.
This checks the coefficients but not the diagrammatic computation behind them.
The relative normalization of the bubble, seagull, and tadpole diagrams is checked by the WT identities: the loop result satisfies the GC identity in full and the LL identity at zero momentum.

The induced action~\labelcref{eq:induced-action} predicts much more: the other two-point functions and the higher-point interactions, up to the operators invisible on vierbein legs.
None of these predictions is guaranteed; each one is a nontrivial check on the induced action.
One such check is already available at the two-point level.
The induced action depends on the LL gauge field $\omega$ through the contorsion~\labelcref{contorsion}, so it also predicts the 1PI two-point functions $\Gamma^{(2)e{\omega}}$ and $\Gamma^{(2){\omega}{\omega}}$.
The latter agrees with Ref.~\cite{Matsuzaki:2020qzf} once one further operator is added, the square~\labelcref{dKdK} of the totally antisymmetrized derivative of the contorsion; \cref{sec:discussion} shows why it is invisible on vierbein legs and fixes its coefficient.

\subsection{Determination from the vierbein two-point function}
\label{sec:determination}

The induced action~\labelcref{eq:induced-action} is established by matching against the 1PI two-point function of \cref{sec:two-point} on the flat torsion-free reference,
\al{
\bar e^\ba{}_\mu\fn{x}
	&=	\delta^\ba_\mu,
&
\bar\omega^\ba{}_{\bb\mu}\fn{x}
	&=	0,
    \label{flat reference background}
}
which also gives $\bar K^\ba{}_{\bb\mu}\fn{x}=0$.
The induced action~\labelcref{eq:induced-action} is a functional of the background alone; its two-point function is reached by shifting the background by the quantum fluctuation,
\al{
\bar e^{\ba}{}_{\mu}\fn{x}
	&\to	\bar e^{\ba}{}_{\mu}\fn{x}+\h e^{\ba}{}_{\mu}\fn{x},
&
\bar\omega^{\ba}{}_{\bb\mu}\fn{x}
	&\to	\bar\omega^{\ba}{}_{\bb\mu}\fn{x}+\h\omega^{\ba}{}_{\bb\mu}\fn{x},
	\label{eq:background shift}
}
and expanding to second order in the fluctuations.
Indices are converted with the background vierbein, and the fluctuation is split into its symmetric and antisymmetric parts,
\al{
\h e_{\mu\nu}\fn{x}
	&:=	\bar e_{\ba\mu}\fn{x}\,\h e^{\ba}{}_{\nu}\fn{x},
&
\h s_{\mu\nu}\fn{x}
	&:=	\h e_{(\mu\nu)}\fn{x},
&
\h a_{\mu\nu}\fn{x}
	&:=	\h e_{[\mu\nu]}\fn{x}.
	\label{eq:ahat}
}
The composite metric then fluctuates by
\al{
\h h_{\mu\nu}\fn{x}
	:=	g_{\mu\nu}\fn{x}-\bar g_{\mu\nu}\fn{x}
	=	2\,\h s_{\mu\nu}\fn{x}
		+\eta_{\ba\bb}\,\h e^{\ba}{}_{\mu}\fn{x}\,\h e^{\bb}{}_{\nu}\fn{x},
	\label{eq:hhat}
}
exactly, terminating at \emph{second} order in the vierbein fluctuation.
Only the symmetric part enters at linear order; the quadratic term is what keeps $\h h$ exactly LL invariant, and it is what carries the antisymmetric part into the metric-sector operators in the matching below.

In the polar variables, \cref{eq:hhat} becomes
\al{
\h h_{\mu\nu}\fn{x}
	&=	\sth_{\mu\nu}\fn{x}
		+\frac14\,\sth_{\mu\rho}\fn{x}\,\sth^{\rho}{}_{\nu}\fn{x},
	\label{eq:stretch-hhat}
}
again exact, and free of the would-be NG boson at every order.
The antisymmetric part $\h a_{\mu\nu}\fn{x}$ is the linearized LL NG mode, $\h a_{\mu\nu}\fn{x}=\h\kappa_{\mu\nu}\fn{x}+\dotsb$ in the expansion of the polar decomposition~\labelcref{eq:polar-decomposition} recorded in \cref{app:polar}.
The exterior derivative of the antisymmetric part,
\al{
\p_{[\lambda}^{}\h a_{\mu\nu]}^{}\fn{x}
\label{eq:da}
}
needs no connection, since the symmetric Christoffels drop out under the total antisymmetrization.
Because $\bar K\fn{x}=0$ on the flat reference, the contorsion is its own quantum fluctuation, $\h K\fn{x}:=K\fn{x}-\bar K\fn{x}=K\fn{x}$ exactly.
Hereafter, a left superscript counts the order in the fluctuations, and the linear part is
\al{
\ord{1}{\h K}_{\ba\bb\mu}\fn{x}
	&=	
		\pn{\h\omega_{\ba\bb\mu}\fn{x}+\os{\bar e}{\mc D}_{\mu}^{}\h a_{\ba\bb}\fn{x}}
		+\pn{
			\os{\bar e}{\mc D}_{\ba}^{}\,\h s_{\bb\bc}\fn{x}
			-\os{\bar e}{\mc D}_{\bb}^{}\,\h s_{\ba\bc}\fn{x}
			}
		\bar e^{\bc}{}_{\mu}\fn{x},
	\label{eq:Khat}
}
where $\os{\bar e}{\mc D}$ is the background-covariant derivative, carrying $\os{\bar g}{\Gamma}$ on spacetime legs and $\os{\bar e}{\Omega}$ on frame legs; unlike the GC-covariant derivative~\labelcref{GC-covariant derivative}, it acts on the LL legs too, and it reduces to $\p$ on the flat reference~\labelcref{flat reference background}.
Frame indices on fluctuations and derivatives are converted with the background vierbein, as in $\h a_{\ba\bb}\fn{x}:=\bar e_{\ba}{}^{\rho}\fn{x}\,\bar e_{\bb}{}^{\sigma}\fn{x}\, \h a_{\rho\sigma}\fn{x}$.
The linearized (true) LL transformation acts as
\al{
\h a_{\ba\bb}\fn{x}
	&\to	\h a_{\ba\bb}\fn{x}+\theta_{\ba\bb}\fn{x},\\
\h\omega_{\ba\bb\mu}\fn{x}
	&\to	\h\omega_{\ba\bb\mu}\fn{x}-\os{\bar e}{\mc D}_{\mu}^{}\theta_{\ba\bb}\fn{x},\\
\h s_{\ba\bb}\fn{x}
	&\to	\h s_{\ba\bb}\fn{x},
}
so each of the two groups in \cref{eq:Khat} is separately invariant. The combination $\h\omega_{\ba\bb\mu}\fn{x}+\os{\bar e}{\mc D}_{\mu}^{}\h a_{\ba\bb}\fn{x}$ is the St\"uckelberg pairing of the connection with the linearized NG mode.
At linear order the six components $\ba\bb$ do not mix, so the pairing reduces to six copies of the abelian St\"uckelberg mechanism; its nonlinear completion is precisely the \emph{hidden local}(-Lorentz) structure of the scenario~\cite{Matsuzaki:2020qzf}.

The two contorsion terms of \cref{eq:induced-action} form the induced torsion sector, whose coefficients are fixed by the minimal coupling of the action~\labelcref{eq:Suv}.
The axial coupling $\eta_1$ in \cref{eq:nonminimal-torsion} would change these two coefficients; the trace-vector coupling $\eta_2$ would add a torsion sector of its own.
In this sector, the antisymmetric vierbein fluctuation enters at linear order through the totally antisymmetric contorsion~\labelcref{eq:Khat}.
Its input to the totally antisymmetric part is the exterior derivative~\labelcref{eq:da} of the linearized NG mode, with \emph{unit} coefficient and nothing else.\footnote{
The coefficient depends on which field one normalizes against.
Our torsion~\labelcref{torsion defined} carries weight $1/2$, as Cartan's does~\cite{Hehl:1976kj}, while the unit-weight torsion of, e.g., Ref.~\cite{Shapiro:2001rz} is twice ours and hence, by \cref{torsion fluctuation}, minus twice the totally antisymmetric contorsion.
Couplings quoted per that torsion are therefore one quarter of the mass and kinetic coefficients in \cref{eq:induced-action}.
The full dictionary, including the convention-dependent signs and the normalization of the axial coupling, will be presented elsewhere~\cite{Oda:2026prep}.
}
The symmetric fluctuation drops out of $\ord{1}{\h K}_{[\lambda\mu\nu]}\fn{x}$ identically.
The composite torsion~\labelcref{torsion defined} carries no further information either,
\al{
\ord{1}{\h{\os{e,\omega}T}}{}_{[\lambda\mu\nu]}\fn{x}=-\ord{1}{\h K}_{[\lambda\mu\nu]}\fn{x},
	\label{torsion fluctuation}
}
identically in both $\h e$ and $\h\omega$, so that ``torsion'' in this sector means, interchangeably, the totally antisymmetric contorsion or the composite torsion.

We now take the second variation of \cref{eq:induced-action} in the vierbein fluctuation, varying the metric sector through \cref{eq:hhat} and the torsion sector through \cref{eq:Khat}.
We require it to reproduce the logarithmic divergences~\labelcref{eq:J1(q)-result} and~\labelcref{eq:J2(q)-result}, which form the pole part of $\Gamma^{(2)ee}|_\tx{spinor}$ in \cref{eq:Gamma-J12}.\footnote{
The sum $J_1+J_2$ of \cref{eq:Gamma-J12} is 1PI by construction, because at one loop every internal line belongs to the single spinor loop and no internal vierbein or LL-gauge-field propagator enters.
A connected vierbein correlator would be a different object, requiring gauge fixing and inversion of the full coupled $(e,\omega)$ Hessian.
}
This requirement determines all five coefficients with no remainder in any tensor structure.
The channel-resolved form of $-\pn{J_{1}+J_{2}}$ is \cref{eq:channel-resolved}.
The divergent coefficient of $\os g R^{2}$ comes out zero,\footnote{
Only the divergent coefficient vanishes.
At zero momentum the $\os gR^{2}$ form factor gives a nonzero finite coefficient, and the irreversible vierbein postulate leaves no bare counterterm that could shift it.
Its value and the full momentum dependence will be reported in Ref.~\cite{Oda:2026prep}.
} and the Gauss--Bonnet combination
\al{
\os g R_{\mu\nu\rho\sigma}\fn{x}\os g R^{\mu\nu\rho\sigma}\fn{x}-4\os g R_{\mu\nu}\fn{x}\os g R^{\mu\nu}\fn{x}+\os g R^{2}\fn{x},
}
topological in $D=4$, is invisible to the two-point function, its second variation vanishing identically.
Only the $m^{4}$ term is non-transverse, exactly as required by the GC WT identity~\labelcref{eq:wi_wi-3}, which ties it to the tadpole $J_{3}$ of \cref{eq:J3(q)-result}.
The LL WT identity~\labelcref{eq:wi_ll-3} at $q=0$ then holds automatically, because $\abb{e}$ is LL invariant.

\subsection{Comparison with the literature}
\label{sec:discussion}

\Cref{eq:induced-action} assembles the ingredients of Sakharov's induced gravity~\cite{Sakharov:1967pk} and extends them to the Einstein--Cartan setting.
The $m^{4}$ term corrects the cosmological constant, and the $\abb{\bar e}\os g R$ term is the induced Einstein--Hilbert action of the composite metric.
The $\os g C^{2}$ term carries the standard one-loop coefficient of a single Dirac spinor~\cite{tHooft:1974toh,Birrell:1982ix}.
In the first-order variables, the symmetric sector of the vierbein thereby acquires a kinetic term of its own.
The two contorsion terms are the two- and four-derivative terms of the LL NG mode, since the vierbein enters the contorsion~\labelcref{eq:Khat} only through the exterior derivative~\labelcref{eq:da}.
The antisymmetric sector of the vierbein thus acquires a kinetic term of its own as well, which completes the vierbein sector of the scenario announced in \cref{sec:intro}.
The torsion sector of \cref{eq:induced-action} lands in the classical arena of quadratic-torsion gravity: the three-parameter teleparallel family of New General Relativity~\cite{Hayashi:1979qx} and its nine-parameter Poincar\'e-gauge extension~\cite{Hayashi:1979wj,Hayashi:1980av,Hayashi:1980ir,Hayashi:1980qp,Hayashi:1980bf}, where these coefficients are free parameters to be constrained by experiment and by the particle spectrum; here, under minimal coupling, they are outputs of a single spinor loop.

As noted in \cref{sec:result}, \cref{eq:induced-action} also predicts the 1PI two-point functions $\Gamma^{(2)e{\omega}}$ and $\Gamma^{(2){\omega}{\omega}}$, because the contorsion~\labelcref{eq:Khat} contains the LL-gauge-field fluctuation $\h\omega$.
The matching in \cref{sec:determination}, however, uses vierbein legs only, and one operator is invisible on vierbein legs,\footnote{
Despite the nearly identical appearance, this is not the last term of \cref{eq:induced-action}.
There the antisymmetrization spans only the three contorsion indices; here it extends over the derivative index as well.
}
\al{
\os{\bar g}\nabla_{[\rho}^{}\bar K_{\lambda\mu\nu]}^{}\fn{x}\,\os{\bar g}\nabla^{[\rho}\bar K^{\lambda\mu\nu]}\fn{x}.
\label{dKdK}
}
On vierbein legs the totally antisymmetric contorsion is the exterior derivative~\labelcref{eq:da}, so each factor of the operator~\labelcref{dKdK} becomes
\al{
\os{\bar g}\nabla_{[\rho}^{}\p_{\lambda}^{}\h a_{\mu\nu]}^{}\fn{x}
	&=	\p_{[\rho}^{}\p_{\lambda}^{}\h a_{\mu\nu]}^{}\fn{x}
	=	0,
}
where the Christoffel symbols drop out under the total antisymmetrization.
The operator~\labelcref{dKdK} can therefore be added to the induced action~\labelcref{eq:induced-action} with an arbitrary coefficient, without changing the vierbein 1PI two-point function.
One vierbein leg is already enough for each factor to vanish, so the operator~\labelcref{dKdK} does not contribute to the mixed 1PI two-point function~$\Gamma^{(2)e{\omega}}$ either; within the displayed dimension-four basis, the induced action~\labelcref{eq:induced-action} fixes that two-point function with nothing left free.

The coefficient of the operator~\labelcref{dKdK} is supplied by the LL-gauge-field 1PI two-point function~$\Gamma^{(2){\omega}{\omega}}$, already computed in Ref.~\cite{Matsuzaki:2020qzf}.
In the result, the mass term and the longitudinal mode carry logarithmic divergences, and the transverse kinetic part carries none.
In the induced action~\labelcref{eq:induced-action}, the contorsion kinetic term produces a divergence in the transverse kinetic part.
At quadratic order about this reference, the operator~\labelcref{dKdK} contributes to the transverse kinetic part and to no other $\omega\omega$ channel.
The two contributions must therefore cancel, and within this basis the cancellation fixes the coefficient of the operator~\labelcref{dKdK}, with no new computation.
The divergences in the mass term and in the longitudinal mode then follow from the induced action~\labelcref{eq:induced-action} with nothing left to tune, and both agree with the result in Ref.~\cite{Matsuzaki:2020qzf}.
This agreement is a nontrivial consistency check.
With the single term~\labelcref{dKdK} added, \cref{eq:induced-action} thus determines the divergent parts of all three 1PI two-point functions $\Gamma^{(2)ee}$, $\Gamma^{(2)e{\omega}}$, and $\Gamma^{(2){\omega}{\omega}}$.
The explicit form of $\Gamma^{(2)e{\omega}}$, the value of the coefficient of the operator~\labelcref{dKdK} in our normalization, the dictionary to the computation of Ref.~\cite{Matsuzaki:2020qzf}, and the manifestly LL-invariant completion on independent $(\h e,\h\omega)$ legs are all deferred to the separate publication~\cite{Oda:2026prep}.

Three lines of work sit adjacent to this result.
The closest is that of Chkareuli and collaborators~\cite{Chkareuli:2011uc,Chkareuli:2017ufz,Chkareuli:2017nzo}, in which the vierbein and spin connection arise as Lorentz (pseudo-)NG modes, reaching Einstein--Cartan in the minimal case; there the NG modes come from \emph{tree-level} spontaneous Lorentz violation through a length-fixing constraint or a tensor vacuum expectation value, whereas everything in \cref{eq:induced-action} is generated radiatively, by a single spinor loop.
Matter-loop effects on Riemann--Cartan backgrounds are well established in the heat-kernel and induced-gravity literature~\cite{Obukhov:1983mm,Grensing:1986tv,Shapiro:2001rz}.
In particular, Chaichian et al.\ derived a cutoff-regulated low-energy Poincar\'e-gauge action from a minimally coupled Dirac field on independent vierbein and Lorentz-connection backgrounds~\cite{Chaichian:2018xdo}, while Ref.~\cite{Nascimento:2021vou} treated topological Nieh--Yan/Pontryagin terms.
There the vierbein and the connection are external backgrounds, and the output is a cutoff-regulated effective action.
Here they are the quantum fluctuations of the irreversible-vierbein scenario, and the output is the off-shell vierbein two-point function together with the WT identities it satisfies.
Finally, the only other off-shell Einstein--Cartan vierbein self-energy known to us is that of Ref.~\cite{Brandt:2024kvs}, computed at one loop in massless pure gravity.
The present one is its fermion-induced, massive counterpart.
Each line of work thus carries one ingredient of the present result: the NG-mode reading of the vierbein, the radiative induction by a Dirac loop, the off-shell vierbein self-energy.
Here, a single massive Dirac spinor produces all three at once; because the action~\labelcref{eq:Suv} is minimally coupled, every coefficient is fixed by the loop.

\section{Summary and discussion}
\label{sec:summary}

In Einstein--Cartan spinor gravity under the irreversible vierbein postulate, only the spinor among the first-order variables has a kinetic term of its own at tree level; those of the vierbein and the LL gauge field come from its quantum fluctuations.
We computed the scheme-independent logarithmically divergent part of the vierbein two-point function at the spinor one-loop level.
The induced kinetic term respects the background gauge invariance and satisfies the WT identities of the true GC and LL invariance, the latter confirmed at zero momentum by the displayed pole terms and at generic momentum for the $1/\epsilon$ pole in \cref{sec:ward-takahashi}.

The pole terms assemble into the local effective action~\labelcref{eq:induced-action} and fix the relative logarithmic contributions of the five displayed channels.
No internal vierbein or LL-gauge-field propagator enters at this order, so the coefficients involve the spinor mass $m$ alone.
For the cosmological and Einstein--Hilbert terms these logarithms correct the bare $\Lambda_\mathrm{cc}$ and $m_\mathrm{P}^{2}$, whereas the Weyl-squared term and the kinetic term of the totally antisymmetric contorsion are new low-energy operators and its mass term adds a torsion-squared coupling independent of $m_\mathrm{P}^{2}$, as explained below \cref{eq:induced-action}.

\Cref{eq:induced-action} fixes only these logarithmic coefficients, not the full low-energy couplings or their ratios.
In a mass-independent scheme they give the one-loop running after minimal subtraction, and with a physical cutoff they are the coefficients of $\ln(\Lambda_\tx{G}^{2}/\mu^{2})$ in \cref{sec:result}; the finite parts differ between minimal subtraction and the cutoff.
A physical cutoff also produces power-sensitive contributions, of order $\Lambda_\tx{G}^{4}$ to the cosmological term and of order $\Lambda_\tx{G}^{2}$ to the Einstein--Hilbert and contorsion-mass terms, with coefficients that depend on the regulator and on the matching.

We have identified the antisymmetric part of the linear vierbein fluctuation as the would-be NG boson of the spontaneously broken LL symmetry, and found that its couplings in the minimally coupled action~\labelcref{eq:Suv} are all fixed by GC and LL gauge invariance.
The polar decomposition~\labelcref{eq:polar-decomposition} removes this mode to all orders: it drops out of the composite metric and can be gauged away entirely.

The result also points beyond the vierbein sector.
Within the displayed dimension-four basis, the contorsion dependence determines the divergent mixed 1PI contribution in the totally antisymmetric channel.
For the LL-gauge-field 1PI two-point function one operator invisible on vierbein legs must be added; once its coefficient is fixed, the result agrees with Ref.~\cite{Matsuzaki:2020qzf}, a nontrivial consistency check discussed in \cref{sec:discussion}.

Whether the induced Weyl-squared term puts an unphysical Ostrogradsky pole into the vierbein propagator cannot be decided from the low-momentum expansion alone; it requires the full coupled $(e,\omega)$ propagator with its nonpolynomial momentum dependence.
The FRG setting for this question is described in \cref{app:frg}.
In the degenerate limit of \cref{sec:degenerate}, the composite metric loses invertibility, and the present description in terms of the metric and the contorsion ceases to apply.
There the present description hands over to the pregeometric phase.

Earlier work in the same model produced the LL-gauge-field kinetic term~\cite{Matsuzaki:2020qzf} and the background spacetime~\cite{Maitiniyazi:2023hts}; a scale-invariant variant produced the Planck scale~\cite{Maitiniyazi:2024zty}.
The present paper adds the vierbein kinetic terms.
Together they fill in the scenario in which both the metric and the torsion of Einstein--Cartan gravity emerge from spinor matter.

The systematic treatment of the mixed and LL-gauge-field sectors, their explicit nonlinear vertices in polar variables, the physical-cutoff matching including its power-sensitive and finite contributions, and the dictionary to the independent-torsion literature will be presented in the separate publication~\cite{Oda:2026prep}.

It would also be intriguing to study the possibility of realizing the structure of the action~\labelcref{eq:induced-action} in terms of non-invertible symmetries~\cite{Kobayashi:2024yqq,Kobayashi:2024cvp}.

\subsection*{Acknowledgement}
We thank Taichiro Kugo for constructive complaints on the presentation of our earlier papers, which prompted us to reorganize the present paper.
The work of S.\,M.\ was supported in part by the Seeds Funding of Jilin University.
The work of Y.\,U.\ was supported in part by the National Natural Science Foundation of China (NSFC) under Grant No.12347112.
The work of K.\,O.\ is supported in part by JSPS KAKENHI Grant Number JP26K00623 and by MEXT SPReAD (AI for Science) Project ID 26279286.
The work of M.\,Y.\ was supported by the NSFC under Grant No.~12205116 and the Seeds Funding of Jilin University.
\appendix
\section*{Appendix}

\section{Review of Einstein--Cartan spinor gravity}
\label{app:review}

This appendix collects the review material supporting \cref{Einstein-Cartan spinor gravity}: the derivations and demonstrations that implement the irreversible vierbein postulate, and the analogy that motivates it.
The material follows Refs.~\cite{Matsuzaki:2020qzf,Maitiniyazi:2023hts} and is included to keep the paper self-contained.

\subsection{The GC transformation and the Lie derivative}
\label{app:review-gc}

Under the GC transformation, the fields transform as
\al{
\label{eq:GCe}
e^\ba{}_\mu\fn{x}& \sr{\tx{GC}}\to e^{\pr\ba}{}_\mu\fn{x'}=e^\ba{}_\nu\fn{x}\pa{M^{-1}}^\nu{}_\mu\fn{x}, \\
\label{eq:GComega}
\omega_\mu\fn{x}&\sr{\tx{GC}}\to \omega'_\mu\fn{x'}=\omega_\nu\fn{x}\pa{M^{-1}}^\nu{}_\mu\fn{x},\\
\label{eq:GCpsi}
\psi\fn{x}&\sr{\tx{GC}}\to\psi'\fn{x'} = \psi\fn{x}
}
where $\pa{M^{-1}}^\nu{}_\mu\fn{x}={\p x^\nu\ov\p x^{\pr\mu}}$ is the inverse of $M^\mu{}_\nu\fn{x}={\p x^{\pr\mu}\ov\p x^\nu}$.
Note that the coordinate differential transforms as
$\df x^\mu
	\sr{\tx{GC}}\to	\df x^{\pr\mu}=M^\mu{}_\nu\fn{x}\df x^\nu$,
so that the volume element transforms as
\al{
\df^{d+1}x
	&\sr{\tx{GC}}\to	\df^{d+1}x'=\pn{\det M}\df^{d+1}x.
}
The matrix $M$ satisfies the GC condition
\al{
\p_{[\lambda} M^\mu{}_{\nu]}\fn{x}=0.
\label{GC condition}
}
Throughout, we employ a notation for a totally antisymmetric tensor
\begin{align}
\mathcal{A}_{\mu\nu}
= \mathcal{A}_{[\mu\nu]}
= \frac{\mathcal{A}_{\mu\nu} - \mathcal{A}_{\nu\mu}}{2} ,
\label{eq:antisym-notation}
\end{align}
and similarly for rank $n$ with weight $1/n!$.

Consequently, the GC transformation is parametrized by the $d+1$ independent functions $x^{\pr\mu}\fn{x}$.
If the condition~\labelcref{GC condition} is not imposed, the transformation is enlarged to $GL(d+1)$, parametrized by the $\pn{d+1}^2$ independent functions $M^\mu{}_\nu\fn{x}$.
Conversely, any $M^\mu{}_\nu\fn{x}$ satisfying the $\frac{d(d+1)^2}{2}$ conditions~\labelcref{GC condition} can locally be written as $M^\mu{}_\nu\fn{x}=\p x^{\pr\mu}/\p x^\nu$ in terms of $d+1$ functions $x^{\pr\mu}\fn{x}$.

The Lie derivative is often introduced in the context of transformations on a spacetime manifold.
It describes how a tensor field changes along the flow generated by a vector field. 
It is closely related to GC transformations, but the two should be distinguished conceptually. 
Let us clarify the difference between a GC transformation and the Lie derivative; see also Appendix~A of Ref.~\cite{Maitiniyazi:2023hts} for a detailed account.
To this end, we first summarize an infinitesimal GC transformation with an infinitesimal vector field $\xi^\mu(x)$:
\al{
x^{\pr\mu}\fn{x}=x^\mu+\xi^\mu\fn{x}.
	\label{infinitesimal GC}
}
Define $\Theta^\mu{}_\nu\fn{x}:=\p_\nu\xi^\mu\fn{x}$.
At first order in $\xi$, the transformation matrices in \cref{eq:GCe,eq:GComega} are $M^\mu{}_\nu\fn{x}=\delta^\mu_\nu+\Theta^\mu{}_\nu\fn{x}$ and $\pa{M^{-1}}^\mu{}_\nu\fn{x}=\delta^\mu_\nu-\Theta^\mu{}_\nu\fn{x}$.
Since partial derivatives commute, $\p_{[\rho}\Theta^\mu{}_{\nu]}\fn{x}=0$, and the GC condition~\labelcref{GC condition} is satisfied.

The infinitesimal transformation~\labelcref{infinitesimal GC} for vierbein, LL gauge field and spinor reads
\al{
\delta_\tx{GC}\,e^\ba{}_\mu\fn{x}
	&=	-e^\ba{}_\nu\fn{x}\Theta^\nu{}_\mu\fn{x}
	=	-\p_\mu\xi^\nu\fn{x}e^\ba{}_\nu\fn{x},
		\label{inifinitesimal GC on e}\\
\delta_\tx{GC}\,\omega^\ba{}_{\bb\mu}\fn{x}
	&=	-\omega^\ba{}_{\bb\nu}\fn{x}\Theta^\nu{}_\mu\fn{x}
	=	-\p_\mu\xi^\nu\fn{x}\omega^\ba{}_{\bb\nu}\fn{x},\\
\delta_\tx{GC}\,\psi\fn{x}
	&=	0.
		\label{inifinitesimal GC on A}
}

A Lie derivative along an infinitesimal vector field $\xi^\mu\fn{x}$ for a field $\Phi(x)$ is defined by comparing the fields at the same coordinate value $x'$ before and after an infinitesimal GC transformation~\labelcref{infinitesimal GC}:
\al{
\mc L_\xi\Phi\fn{x}
:= \Phi\fn{x'}-\Phi'\fn{x'}
&= \Phi\Fn{x+\xi\fn{x}}-\Phi'\fn{x'}\nn
&= \Pn{\Phi\fn{x}+\xi\fn{x}\Phi\fn{x}}-\Phi'\fn{x'}\nn
&= \xi\fn{x}\Phi\fn{x}+\pn{\Phi\fn{x}-\Phi'\fn{x'}}\nn
&= \xi\fn{x}\Phi\fn{x}-\delta_\tx{GC}\Phi\fn{x},
\label{Lie derivative given}
}
where $\xi\fn{x}$ appearing as an argument of $\Phi$ in the first line denotes the set of $\pn{d+1}$ components $\pn{\xi^0\fn{x},\dots,\xi^d\fn{x}}$.
Recall that $\xi\fn{x}$ appearing elsewhere is the differential operator $\xi^\mu\fn{x}\p_\mu$ in \cref{sec:LD}.
The GC transformation can be expressed in terms of the Lie derivative as
\al{
\delta_\tx{GC}\,\Phi\fn{x}
	&=	\xi\fn{x}\Phi\fn{x}-\mc L_\xi\Phi\fn{x}.
 \label{app:eq: GC trans}
}
We may therefore define the LD transformation, introduced in \cref{sec:LD}, as
\al{
\delta_\tx{LD}\,\Phi\fn{x}
	&:=	\delta_\tx{GC}\,\Phi\fn{x}-\xi\fn{x}\Phi\fn{x}
	=	-\mc L_\xi\,\Phi\fn{x}.
 \label{app: eq: LD trans}
}

An advantage of the GC transformation over the LD transformation is that the former commutes with gauge transformations, whereas the latter generally does not~\cite{Daum:2009dn,Daum:2010bc,Daum:2010qt,Daum:2013fu}.
This difference originates from the fact that the Lie derivative compares a field at different spacetime points, which are affected differently by gauge transformations.

Nevertheless, the GC WT identity in the main text is derived with the LD transformation rather than the GC one.
It is an identity for the effective action as a functional of the fields on a fixed coordinate chart, and on a fixed chart the GC invariance of the action becomes precisely the invariance under $\delta_\tx{LD}\Phi=-\mc L_\xi\Phi$.
The GC invariance itself states that the action is unchanged when the integration variable is relabeled $x\to x^\pr$ together with the fields $\Phi\to\Phi^\pr$, the Jacobian of the measure canceling that of the transformed integrand.
In the fixed-argument variation $\delta_\tx{GC}$, the integration variable is untouched and no measure Jacobian arises, so $\delta_\tx{GC}$ alone is not a symmetry.
For the cosmological-constant term, $\delta_\tx{GC}\abb{e}=-\abb{e}\,\p_\mu\xi^\mu$ is not a total derivative, while $\delta_\tx{LD}\abb{e}=-\p_\mu\pn{\xi^\mu\abb{e}}$ is.

\subsection{The GC gauge field and the Levi-Civita connection}
\label{app:review-upsilon}

The gauge field associated with the GC transformation is denoted by $\Upsilon_\mu\fn{x}$ in the matrix notation $\mt{\Upsilon_\mu\fn{x}}^\alpha{}_\beta:=\Upsilon^\alpha{}_{\beta\mu}\fn{x}$ and transforms as
\al{
\label{GCgaugetrans}
\Upsilon_\mu\fn{x}
	&\sr{\tx{GC}}\to	\Upsilon'_\mu\fn{x'}
	=	\mt{ M\fn{x}\Upsilon_\nu\fn{x}M^{-1}\fn{x} -\Paren{\p_\nu M\fn{x}}M^{-1}\fn{x} }\mt{M^{-1}\fn{x}}^\nu{}_\mu.
}
This transformation law resembles the LL transformation law~\labelcref{eq:LLgaugetrans} of $\omega_\mu\fn{x}$ in the adjoint representation. However, the GC transformation of $\Upsilon_\mu\fn{x}$ contains an additional factor of $M^{-1}$, which also transforms the spacetime index.

For the minimal GC gauging reviewed here, we choose $\Upsilon_\mu\fn{x}$ to be torsion free, so that
\al{
\Upsilon^\alpha{}_{\beta\mu}\fn{x}
	&=	\Upsilon^\alpha{}_{(\beta\mu)}\fn{x},
		\label{separation of Gamma}
}
where we introduced a notation for a totally symmetric tensor
\al{
\mathcal{S}_{\mu\nu}
= \mathcal{S}_{(\mu\nu)}
= \frac{\mathcal{S}_{\mu\nu} + \mathcal{S}_{\nu\mu}}{2} .
\label{eq:sym-notation}
}
The symmetric part $\Upsilon^\alpha{}_{(\beta\mu)}\fn{x}$ has $\frac{(d+1)^2(d+2)}{2}$ degrees of freedom out of the $\pn{d+1}^3$ of a general $\Upsilon^\alpha{}_{\beta\mu}\fn{x}$.\footnote{
For a general affine connection one may retain $\Upsilon^\alpha{}_{[\beta\mu]}$, with $\Upsilon^\alpha{}_{\beta\mu}=\Upsilon^\alpha{}_{(\beta\mu)} + \Upsilon^\alpha{}_{[\beta\mu]}$.
This antisymmetric part has $\frac{d(d+1)^2}{2}$ degrees of freedom and is the spacetime-torsion tensor; neither GC covariance nor the GC condition~\labelcref{GC condition} forces it to vanish.
Under the GC condition, the inhomogeneous term in \cref{GCgaugetrans} is symmetric in $\beta,\mu$, so $\Upsilon^\alpha{}_{[\beta\mu]}$ transforms homogeneously as a tensorial torsion component.
It is not required for gauging GC and is set to zero in the present construction; torsion in the dynamical theory is carried by the independent LL sector.
}

We also impose the metricity condition on $\Upsilon$:
\al{
\os \Upsilon\nabla_\alpha g_{\beta\mu}\fn{x}= \Paren{\p_\alpha+ \Upsilon_\alpha\fn{x}} g_{\beta\mu}\fn{x}
=	\p_\alpha g_{\beta\mu}\fn{x} 
	-  g_{\rho\mu}\fn{x} \Upsilon^{\rho}{}_{\beta\alpha}\fn{x}
	- g_{\beta\rho}\fn{x} \Upsilon^\rho{}_{\mu\alpha}\fn{x}
=0.
}
Together with the torsion-free choice~\labelcref{separation of Gamma}, the metricity condition uniquely identifies the GC gauge field with the Levi--Civita connection~\labelcref{eq:Levi-Civita}:
\al{
\Upsilon^\alpha{}_{(\beta\mu)}\fn{x}
=\os g\Gamma{}^\alpha{}_{\beta\mu}\fn{x}.
 \label{Levi-Civita given}
}

Under the infinitesimal GC transformation~\labelcref{infinitesimal GC}, the GC-gauge field transforms as
\al{
\delta_\tx{GC}\,\Upsilon^\alpha{}_{\beta\mu}\fn{x}
	&=	\Theta^\alpha{}_\gamma\fn{x}\Upsilon^\gamma{}_{\beta\mu}\fn{x}
		-\Upsilon^\alpha{}_{\delta\mu}\fn{x}\Theta^\delta{}_\beta\fn{x}
		-\Upsilon^\alpha{}_{\beta\nu}\fn{x}\Theta^\nu{}_\mu\fn{x}
		-\p_\mu\Theta^\alpha{}_\beta\fn{x}.
}

\subsection{Demonstrations of the degenerate limit}
\label{app:review-demos}

Let us here demonstrate the degenerate limit in $D=4$ more specifically.
A first key point is that the invariant spacetime measure always includes the determinant of the vierbein:
\al{
\abb{e\fn{x}}:=\det_{\ba,\mu}e^\ba{}_\mu\fn{x}
=	\frac{1}{4!0!}
\ep{\ba\bb\bc\bd}\ep{\mu\nu\rho\sigma}
e^\mathbf{a}{}_\mu\fn{x}
e^\mathbf{b}{}_\nu\fn{x}
e^\mathbf{c}{}_\rho\fn{x}
e^\mathbf{d}{}_\sigma\fn{x} 
=	\lambda_0\fn{x}\cdots\lambda_3\fn{x}.
	\label{eq: product of eigenvalues of vielbein}
}
Here, we introduced the Levi-Civita symbol
\al{
\ep{\mu_0\dots\mu_3}
	&=	\begin{cases}
		1	&	\tx{when $\pn{\mu_0,\dots,\mu_3}$ is even permutation of $\pn{0,\dots,3}$,}\\
		-1	&	\tx{when $\pn{\mu_0,\dots,\mu_3}$ is odd permutation of $\pn{0,\dots,3}$,}\\
		0	&	\tx{otherwise,}
		\end{cases}
}
and
\al{
\ep{\bs{a_0}\dots\bs{a_3}}
	&=	\begin{cases}
		1	&	\tx{when $\pn{\bs{a_0},\dots,\bs{a_3}}$ is even permutation of $\pn{\bs0,\dots,\bs 3}$,}\\
		-1	&	\tx{when $\pn{\bs{a_0},\dots,\bs{a_3}}$ is odd permutation of $\pn{\bs0,\dots,\bs 3}$,}\\
		0	&	\tx{otherwise.}
		\end{cases}
}
It is apparent that $\abb{e\fn{x}}\to 0$ in the limit that some of the $\lambda_a\fn{x}$ are taken to zero. 
This fact implies that the cosmological constant term
\al{
\int \df^4 x\abb{e\fn{x}} \Lambda_\tx{cc}
\label{eq: cosmilogical constant term}
}
is compatible with the degenerate limit.

The determinant formula~\labelcref{eq: product of eigenvalues of vielbein} is extended to
\al{
\abb{e\fn{x}}e_\ba{}^\mu\fn{x}
	&=	{1\ov3!1!}\ep{\ba\bb\bc\bd}\ep{\mu\nu\rho\sigma}e^\bb{}_\nu\fn{x}e^\bc{}_\rho\fn{x}e^\bd{}_\sigma\fn{x},
 \label{eq: determinant e}
 \\
\abb{e\fn{x}}e_{[\ba}{}^\mu\fn{x}e_{\bb]}{}^\nu\fn{x}
	&=	{1\ov2! 2!}\ep{\ba\bb\bc\bd}\ep{\mu\nu\rho\sigma}e^\bc{}_\rho\fn{x}e^\bd{}_\sigma\fn{x},
	\label{eq: determinant ee}
	\\
\abb{e\fn{x}}e_{[\ba}{}^\mu\fn{x}e_\bb{}^\nu\fn{x}e_{\bc]}{}^\rho\fn{x}
	&=	{1\ov1!3!}\ep{\ba\bb\bc\bd}\ep{\mu\nu\rho\sigma}e^\bd{}_\sigma\fn{x},
	\label{eq: determinant eee}
	\\
\abb{e\fn{x}}e_{[\ba}{}^\mu\fn{x}e_\bb{}^\nu\fn{x}e_\bc{}^\rho\fn{x}e_{\bd]}{}^\sigma\fn{x}
	&=	{1\ov0!4!}\ep{\ba\bb\bc\bd}\ep{\mu\nu\rho\sigma}.
	\label{eq: determinant eeee}
}
Multiplied by the determinant, antisymmetrized products of inverse vierbeins are thus polynomial in the vierbeins.
Inverse vierbeins are therefore not prohibited altogether: in $D=4$, the left-hand sides of \cref{eq: determinant e,eq: determinant ee,eq: determinant eee,eq: determinant eeee} are the only combinations compatible with the degenerate limit.

The kinetic term of the spinor field contains the combination of \cref{eq: determinant e}:
\al{
S
&= -\frac{1}{2}\int \df^4x \abb{e\fn{x}} 
\pn{
\ol\psi\fn{x} 
e_\mathbf{a}{}^\mu\fn{x}
\gamma^\mathbf{a}
\os\omega{\mathcal D}_\mu\psi\fn{x}
-
\Pn{\os\omega{\mathcal D}_\mu\ol\psi\fn{x}}
e_\mathbf{a}{}^\mu\fn{x} 
\gamma^\mathbf{a}
\psi\fn{x} 
}\nn
&= -\frac{1}{2}\int \df^4x \abb{e\fn{x}}e_\mathbf{a}{}^\mu\fn{x}
\pn{
\ol\psi\fn{x} 
\gamma^\mathbf{a}
\os\omega{\mathcal D}_\mu\psi\fn{x}
-
\Pn{\os\omega{\mathcal D}_\mu\ol\psi\fn{x}}
\gamma^\mathbf{a}
\psi\fn{x} 
}.
\label{eq: kinetic term of spinor}
}
Therefore, this term is suitable in the degenerate limit.
Note that the mass term of the spinor field is also accepted for the same reason as the cosmological constant term.

Unlike an ordinary Yang--Mills theory, the linear term in the field strength tensor of the LL gauge field $\os\omega{\mc F}^{\ba\bb}{}_{\mu\nu}\fn{x}$ is formed as an invariant term together with inverse vierbeins:
\al{
S=\frac{m_\mathrm{P}^2}{2}\int \df^4 x\abb{e\fn{x}} e_{[\ba}{}^\mu\fn{x} e_{\bb]}{}^\nu\fn{x} \os\omega{\mc F}^{\ba\bb}{}_{\mu\nu}\fn{x}.
\label{eq: linear in F}
}
It does not diverge in the degenerate limit because \cref{eq: determinant ee} is proportional to
\al{
\Pn{\diag\fn{\lambda_0,\lambda_1,\lambda_2,\lambda_3}}^{\otimes 2}.
}
Hereafter we leave the $x$ dependence of $\lambda_a$, $N$, and $M$ implicit when apparent.

Let us now demonstrate several cases that are not compatible with the irreversible vierbein postulate.
First, we consider the case of $\abb{e\fn{x}}g^{\mu\nu}\fn{x}$ where the inverse metric field is
\al{
g^{\mu\nu}\fn{x}=e_{(\ba}{}^\mu\fn{x} e_{\bb)}{}^\nu\fn{x}\eta^{\ba\bb}
&= M\diag\fn{\lambda_0^{-1},\dots,\lambda_3^{-1}}N^{-1}\,\eta\,\pn{N^{-1}}^\t\diag\fn{\lambda_0^{-1},\dots,\lambda_3^{-1}}M^\t\nn
&= M\diag\fn{-\lambda_0^{-2},\lambda_1^{-2},\lambda_2^{-2},\lambda_3^{-2}}M^\t,
}
where we used $N^{-1}\eta\pn{N^{-1}}^\t=\eta$.
Note that it is different from $\abb{e\fn{x}}e_{[\ba}{}^\mu\fn{x} e_{\bb]}{}^\nu\fn{x}$ which vanishes for the contraction with $\eta^{\ba\bb}$.
Then, we find
\al{
\abb{e\fn{x}}g^{\mu\nu}\fn{x}&\propto (\lambda_0\cdots\lambda_3)\diag\fn{-\lambda_0^{-2},\lambda_1^{-2},\lambda_2^{-2},\lambda_3^{-2}}\nn
&\quad=\diag\fn{-{\lambda_1\lambda_2\lambda_3\ov\lambda_0},{\lambda_0\lambda_2\lambda_3\ov\lambda_1},{\lambda_0\lambda_1\lambda_3\ov\lambda_2},{\lambda_0\lambda_1 \lambda_2\ov\lambda_3}}.
    \label{inverse metric counting}
}
This contains the inverse of $\lambda_a$, so that terms with this combination are prohibited as being divergent for $\lambda_a\to 0$.
Its typical case is the kinetic term of scalar fields 
\al{
S_\phi&=\int\df^4x\abb{e\fn{x}}\sqbr{-{1\ov2}g^{\mu\nu}\fn{x}\paren{\p_\mu\phi\fn{x}}\paren{\p_\nu\phi\fn{x}}}
\nn&= \int\df^4x\abb{e\fn{x}}g^{\mu\nu}\fn{x}\sqbr{-{1\ov2}\paren{\p_\mu\phi\fn{x}}\paren{\p_\nu\phi\fn{x}}}.
}
Thus, the kinetic term of scalar fields does not meet the irreversible vierbein postulate.

Gauge fields are also not allowed to have their kinetic term
\al{
S_\tx{Gauge}&=\int\df^4x\abb{e\fn{x}}\sqbr{\pm {1\ov2}g^{\mu\mu'}\fn{x}g^{\nu\nu'}\fn{x}\tr [F_{\mu\nu}\fn{x}F_{\mu'\nu'}\fn{x}]}
\nn&=\int\df^4x\abb{e\fn{x}}g^{\mu\mu'}\fn{x}g^{\nu\nu'}\fn{x} \sqbr{\pm {1\ov2}\tr [F_{\mu\nu}\fn{x}F_{\mu'\nu'}\fn{x}]},
}
where the minus sign is for ordinary gauge theory, while the plus sign is for the LL gauge theory.
There is a divergent combination for $\lambda_a\to 0$:
\al{
\abb{e\fn{x}}g^{\mu\mu'}\fn{x}g^{\nu\nu'}\fn{x}&\propto (\lambda_0\cdots\lambda_3) [\diag\fn{-\lambda_0^{-2},\lambda_1^{-2},\lambda_2^{-2},\lambda_3^{-2}}]^{\otimes 2}.
\label{eq: dete gg 2}
}
In the same manner, we see that the symmetric and antisymmetric kinetic terms for the vierbein 
\al{
S_{e,\tx{sym}}&=\int\df^4x\abb{e\fn{x}}\sqbr{-{Z_{es}\ov2}g^{\mu\mu'}\fn{x}g^{\nu\nu'}\fn{x}\eta_{\ba\bb}\paren{\p_{(\mu} e^\ba{}_{\nu)}\fn{x}}\paren{\p_{(\mu'} e^\bb{}_{\nu')}\fn{x}}},\\
S_{e,\tx{antisym}}&=\int\df^4x\abb{e\fn{x}}\sqbr{-{Z_{ea}\ov2}g^{\mu\mu'}\fn{x}g^{\nu\nu'}\fn{x}\eta_{\ba\bb}\,\os{e,\omega}{T}{}^\ba{}_{\mu\nu}\fn{x}\os{e,\omega}{T}{}^\bb{}_{\mu'\nu'}\fn{x}},
}
are forbidden in the degenerate limit, where the torsion $\os{e,\omega}{T}{}^\ba{}_{\mu\nu}\fn{x}$ is defined in \cref{torsion defined}.
The antisymmetric term is precisely of the type taken as the fundamental gravitational Lagrangian in New General Relativity~\cite{Hayashi:1979qx}, where it belongs to the three-parameter family of torsion-quadratic Lagrangians.

Finally, we consider the Einstein--Hilbert action consisting solely of vierbeins,
\al{
S_\tx{EH}=\frac{m_\mathrm{P}^2}{2}\int\df^4 x\abb{e\fn{x}}\os g R\fn{x},
}
where $\os g R$ is the Ricci scalar.
Here, the Riemann tensor, the Ricci tensor and the Ricci scalar are defined respectively as
\al{
\os g  R^\mu{}_{\nu\rho\sigma}\fn{x}
&=	\p_\rho \os g\Gamma^\mu{}_{\nu\sigma}\fn{x}-\p_\sigma \os g\Gamma^\mu{}_{\nu\rho}\fn{x}+\os g\Gamma^\mu{}_{\lambda\rho}\fn{x}\os g\Gamma^\lambda{}_{\nu\sigma}\fn{x}-\os g\Gamma^\mu{}_{\lambda\sigma}\fn{x}\os g\Gamma^\lambda{}_{\nu\rho}\fn{x},\\ 
\os g  R_{\nu\sigma}\fn{x}
&= g^{\mu\rho}\fn{x}\os g   R_{\mu\nu\rho\sigma}\fn{x}\nn
&=\p_\mu\os g\Gamma^\mu{}_{\nu\sigma}\fn{x}-\p_\sigma\os g\Gamma^\mu{}_{\nu\mu}\fn{x}+\os g\Gamma^\mu{}_{\lambda\mu}\fn{x}\os g\Gamma^\lambda{}_{\nu\sigma}\fn{x}-\os g\Gamma^\mu{}_{\lambda\sigma}\fn{x}\os g\Gamma^\lambda{}_{\nu\mu}\fn{x},\\
\os g R\fn{x}
&=g^{\nu\sigma}\fn{x}\os g  R_{\nu\sigma}\fn{x}\nn
&=g^{\nu\sigma}\fn{x}\paren{\p_\mu\os g\Gamma^\mu{}_{\nu\sigma}\fn{x}-\p_\sigma\os g\Gamma^\mu{}_{\nu\mu}\fn{x}+\os g\Gamma^\mu{}_{\lambda\mu}\fn{x}\os g\Gamma^\lambda{}_{\nu\sigma}\fn{x}-\os g\Gamma^\mu{}_{\lambda\sigma}\fn{x}\os g\Gamma^\lambda{}_{\nu\mu}\fn{x}},
\label{eq:ricci-convention}
}
where the Levi-Civita connection $\os g \Gamma^\mu{}_{\alpha\beta}\fn{x}$ is defined in \cref{eq:Levi-Civita}.
Since the Levi-Civita connection contains one factor of the inverse metric, the $\os g\Gamma\,\os g\Gamma$ terms in \cref{eq:ricci-convention} come with two such factors, and the further contraction with $g^{\nu\sigma}\fn{x}$ adds no net factor in the leading piece.\footnote{In $\os g\Gamma^\mu{}_{\lambda\sigma}\fn{x}\os g\Gamma^\lambda{}_{\nu\mu}\fn{x}$ the derivatives carry $\mu$ and $\lambda$, while $\nu$ and $\sigma$ sit on the metrics being differentiated; contracting them with $g^{\nu\sigma}\fn{x}$ therefore costs no further inverse metric.}
The derivative structures in $\abb{e\fn{x}}\os gR\fn{x}$ thus come with the divergent coefficient combination~\labelcref{eq: dete gg 2}, and $S_\tx{EH}$ does not meet the irreversible vierbein postulate.\footnote{The power counting gives an upper bound on the rate of divergence: in explicit degenerate families cancellations lower the actual rate, but $\abb{e\fn{x}}\os gR\fn{x}$ still diverges, and only that matters for the postulate.}

\subsection{Topological terms}
\label{app:review-topological}

We note here that the use of \cref{eq: determinant eeee} allows us to write the following terms:
\al{
S_\text{Euler}
	&=	{1\ov32}\int\df^4 x\,\ep{\mu\nu\rho\sigma}\ep{\ba\bb\bc\bd} \os\omega{\mc F}^{\ba\bb}{}_{\mu\nu}\fn{x}\os\omega{\mc F}^{\bc\bd}{}_{\rho\sigma}\fn{x},
}
which is the Euler term.
In $D=4$ it is a total derivative even with torsion, i.e., for an independent Lorentz gauge field; this is the extended Bach--Lanczos identity~\cite{Hayashi:1980bf}.
In addition, using the Levi-Civita symbol $\ep{\mu\nu\rho\sigma}$ and the tangent-space metric $\eta_{\ba\bb}$, we can write down the following terms:
\al{
S_\text{Immirzi}
 &=	{m_\mathrm{P}^{2}\ov\gamma}\int\df^4x \ep{\mu\nu\rho\sigma} \eta_{\ba\bc} \eta_{\bb\bd}{1\ov4}
		\os\omega{\mc F}^{\ba\bb}{}_{\mu\nu}\fn{x}e^\bc{}_\rho\fn{x}e^\bd{}_\sigma\fn{x},\\
S_\text{Nieh-Yan}
	&=	\int\df^4x\ep{\mu\nu\rho\sigma}\pn{
			\eta_{\ba\bb}
			\os{e,\omega}{T}{}^\ba{}_{\mu\nu}\fn{x}\os{e,\omega}{T}{}^\bb{}_{\rho\sigma}\fn{x}
			-{1\ov2} \eta_{\ba\bc} \eta_{\bb\bd}\os\omega{\mc F}^{\ba\bb}{}_{\mu\nu}\fn{x}e^\bc{}_\rho\fn{x}e^\bd{}_\sigma\fn{x}
			},
			\label{Nieh-Yan action}\\
S_\text{Pontryagin}
		&=	\int\df^4x\ep{\mu\nu\rho\sigma}\eta_{\ba\bc} \eta_{\bb\bd}{1\ov8}\os\omega{\mc F}^{\ba\bb}{}_{\mu\nu}\fn{x}\os\omega{\mc F}^{\bc\bd}{}_{\rho\sigma}\fn{x},
}
Of these terms, the Nieh--Yan and Pontryagin terms, together with the Euler term above, are total derivatives.
The Immirzi term is not: it is $m_\mathrm{P}^{2}/(2\gamma)$ times the difference between the torsion-squared combination and the Nieh--Yan invariant.
It modifies the one-derivative $e\omega$ mixing and the algebraic $\omega\omega$ term, while on vierbein legs it is a total derivative at quadratic order and does not enter the $ee$ two-point function computed here.
Its coefficient is written with the Immirzi parameter $\gamma$: Together with the Einstein--Cartan term of \cref{eq:Suv}, it gives the Holst form $\frac{m_\mathrm{P}^{2}}{2}\,e_\ba{}^{\mu}e_\bb{}^{\nu}\bigl(\os\omega{\mc F}^{\ba\bb}{}_{\mu\nu}-\frac{1}{2\gamma}\,\eta^{\ba\ba'}\eta^{\bb\bb'}\ep{\ba'\bb'\bc\bd}\os\omega{\mc F}^{\bc\bd}{}_{\mu\nu}\bigr)$ of Refs.~\cite{Perez:2005pm,Freidel:2005sn}.
None of the four supplies a kinetic term of its own for the vierbein or the LL gauge field, and we do not include them in the action~\labelcref{eq:Suv}; for the Immirzi term this is the choice $1/\gamma=0$ about the bare action, not a consequence of topology.

What the Immirzi term would do is easy to state in our variables.
Its torsion-squared part is parity odd: It couples the antisymmetrized contorsion~$K_{[\lambda\mu\nu]}$ to the trace vector $K^{\lambda}{}_{\mu\lambda}$.
Only $K_{[\lambda\mu\nu]}$ enters the minimally coupled spinor action, so at one loop the trace vector has no kinetic term and stays auxiliary.
Eliminating the trace vector multiplies the tree-level mass term of $K_{[\lambda\mu\nu]}$ by $1+1/\gamma^{2}$.
In the standard treatment $K_{[\lambda\mu\nu]}$ is auxiliary as well, and eliminating it gives the four-fermion contact interaction of Refs.~\cite{Perez:2005pm,Freidel:2005sn}, whose coefficient is inversely proportional to this mass term and hence carries the factor $\gamma^{2}/(\gamma^{2}+1)$.
Here $K_{[\lambda\mu\nu]}$ propagates, with the kinetic term induced in \cref{eq:induced-action}, and the contact interaction is only the low-momentum limit of its exchange.

\subsection{Motivation: analogy with electroweak symmetry breaking}
\label{app:review-ewsb}

The motivation for imposing the irreversible vierbein postulate is to define the symmetric vacuum of the LL gauge symmetry.
The vierbein belongs to the fundamental representation of the LL symmetry, and its vacuum expectation value spontaneously breaks the LL symmetry.
This is analogous to the Higgs doublet field $H\fn{x}$ in the Standard Model. The doublet transforms in the fundamental representation of the local $SU(2)_L$ gauge symmetry and spontaneously breaks $SU(2)_L$ by acquiring a vacuum expectation value.
In the Standard Model Effective Field Theory (SMEFT), inverse powers of the Higgs norm are generally excluded.
The operator $(H^\dagger\fn{x} H\fn{x})^{-1}$ is gauge invariant but nonanalytic in the symmetric phase, where $v := \sqrt{2}\Braket{\ab{H\fn{x}} } = 0$.\footnote{
For example, we do not assume $e^{-g/(H^\dagger H)}$ even though it is finite at the symmetric phase.
}

Even if we allow a singularity in the vacuum, this inverse norm remains incompatible with the cutoff argument below.
In the unitary gauge it is
\al{
(H^\dagger\fn{x} H\fn{x})^{-1}
&= \frac{2}{(v+h\fn{x})^2}\nn
&= \frac{2}{v^2}
- \frac{4}{v^2} \left(\frac{h\fn{x}}{v}\right)
+ \frac{6}{v^2}\left(\frac{h\fn{x}}{v}\right)^2
- \frac{8}{v^2}\left(\frac{h\fn{x}}{v}\right)^3
+ \cdots,
\label{eq:expansionHiggs}
}
which generates the infinite number of the higher-dimensional operators all suppressed by powers of $v$.
This suggests the existence of heavy degrees of freedom whose masses are proportional to $v$, indicating the low cutoff scale \cite{Falkowski:2019tft}.
Such a scenario is incompatible with the SMEFT, whose validity requires the cutoff scale to be well above the electroweak scale.

The expansion of the inverse vierbein~\labelcref{eq:expansionInversevierbein} is analogous to the expansion of the Higgs field~\labelcref{eq:expansionHiggs}. In some extensions of the Standard Model, a nonvanishing electroweak vacuum expectation value $v$ is generated dynamically through quantum effects, e.g., via the Coleman--Weinberg mechanism~\cite{Coleman:1973jx,Gildener:1976ih,Bardeen:1995kv,Hempfling:1996ht,Meissner:2006zh,Foot:2007as,Iso:2009ss,Hamada:2012bp,Iso:2012jn,Hambye:2013dgv,Hamada:2020wjh,Hamada:2021jls} or in technicolor-like~\cite{Hill:2002ap,Sannino:2009za,Cacciapaglia:2020kgq}, hypercolor-scalegenesis~\cite{Hur:2011sv,Holthausen:2013ota,Kubo:2015cna,Kubo:2016kpb,Kubo:2017wbv,Ouyang:2018eub,Haba:2005jq,Antipin:2014qva,Haba:2015lka,Haba:2015qbz,Ishida:2016ogu,Ishida:2016fbp,Haba:2017wwn,Haba:2017quk,Ishida:2017ehu,Ishida:2019gri}, little-Higgs~\cite{Schmaltz:2005ky}, and extra-dimensional~\cite{Hosotani:2025ala} scenarios.
Motivated by this analogy, we may consider the possibility that a nondegenerate vierbein background satisfying $\abb{\bar e\fn{x}}\neq0$ is likewise generated dynamically in our gravitational model. In our previous works, we demonstrated that such a nondegenerate vierbein background can indeed emerge from spinor-field dynamics~\cite{Maitiniyazi:2023hts,Maitiniyazi:2024zty}.

Furthermore, we note that degenerate vierbein configurations arise in spacetime topology-changing processes and are expected to play an important role in quantum gravity~\cite{Tseytlin:1981ks,Horowitz:1990qb}.

\section{Functional renormalization group}
\label{app:frg}

The FRG is a powerful method to investigate the nonperturbative quantum dynamics of the theory. 
In the FRG, quantum fluctuations are integrated out successively according to their momentum scale, resulting in a scale-dependent 1PI effective action $\Gamma_k$.
The infrared (IR) cutoff scale $k$ suppresses fluctuations with momenta $p^2\lesssim k^2$, while modes with $p^2\gtrsim k^2$ are already integrated out. 
Consequently, $\Gamma_k$ interpolates between the microscopic action $S$ in the UV and the full quantum effective action $\Gamma$ in the IR,
\begin{align}
\Gamma_{k\to\Lambda} &\simeq S,&
\Gamma_{k\to0} &= \Gamma,
\end{align}
where $\Lambda$ denotes the UV cutoff scale.
For our gravitational model, the initial action is identified with \cref{eq:Suv} at $k=\Lambda_\tx{G}$.
The evolution of $\Gamma_k$ with respect to the renormalization-group scale $k$ is governed by the Wetterich equation \cite{Wetterich:1992yh} (see also \cite{Morris:1993qb,Ellwanger:1993mw,Reuter:1993kw}),
\begin{align}
k\partial_k \Gamma_k
=
\frac{1}{2}
\mathrm{STr}
\fnl{
\pn{\Gamma_k^{(2)}+\mathcal R_k}^{-1}
k\partial_k\mathcal R_k
}.
\end{align}
Here, $\Gamma_k^{(2)}$ denotes the second functional derivative of $\Gamma_k$ with respect to the fields, corresponding to the full inverse propagator, while $\mathcal R_k$ is the IR regulator that suppresses low-momentum fluctuations. The supertrace $\mathrm{STr}$ represents a sum over all field components together with an integration over momenta, where fermionic contributions enter with a minus sign. Owing to the insertion of $k\partial_k\mathcal R_k$, the flow equation receives dominant contributions only from modes with momenta around $p^2\sim k^2$, allowing for a continuous integration of quantum fluctuations from the UV to the IR.

In our gravitational model, we infer that the scale-dependent effective action takes the form
\al{
\Gamma_k
=
\Gamma_k\fnl{\bar\Phi}
+
\frac{1}{2}\int \df^4x \abb{\bar e\fn{x}}
\left[
\hat e^\bb{}_\nu\fn{x}
\Paren{\Gamma_{k}^{(2)ee}[\p;\bar\Phi]}_{\bb\bc}{}^{\nu\rho}
\hat e^\bc{}_\rho\fn{x}
\right]
+\cdots,
}
where $\bar\Phi=(\bar e^\ba{}_\mu\fn{x},\bar\omega^\ba{}_{\bb\mu}\fn{x})$ denotes the background vierbein and LL gauge field, $\Gamma_k[\bar\Phi]$ is the effective action depending solely on the background fields, the term linear in $\hat e$ is omitted because it does not enter the propagator, and the ellipsis represents higher-order vertices of the vierbein fluctuation $\hat e$, the LL gauge field sector, and the spinor sector.

The full two-point function of the vierbein in momentum space is defined by
\al{
\Paren{\Gamma_{k}^{(2)ee}[q;\bar\Phi]}_{\bb\bc}{}^{\nu\rho}
=
\frac{\delta^2\Gamma_k}
{\delta e^\bb{}_\nu(-q)\,
\delta e^\bc{}_\rho(q)}\bigg|_{\Phi=\bar\Phi}.
\label{eq:Gamma-ee-k}
}
This is the $ee$ block of the Hessian $\Gamma_k^{(2)}$, not a propagator.
The propagator is the inverse of the full Hessian, and its vierbein block differs from the inverse of \cref{eq:Gamma-ee-k} whenever $\hat e$ mixes with $\hat\omega$.

As shown in \cref{sec:two-point}, the spinor one-loop contribution induces
\al{
\Paren{\Gamma_{k=0}^{(2)ee}[q;\bar\Phi]}_{\bb\bc}{}^{\nu\rho}
\bigg|_\tx{spinor}
=
J_1(q)_{\bb\bc}{}^{\nu\rho}
+
J_2(q)_{\bb\bc}{}^{\nu\rho},
}
where $J_1(q)$ and $J_2(q)$ are given in \cref{eq:J1(q)-result,eq:J2(q)-result} with $1/\epsilon\sim\ln \Lambda_\tx{G}$, respectively.
That is, the $ee$ block $\Gamma_{k=0}^{(2)ee}[q;\bar\Phi]$ is the two-point function $\Gamma^{(2)ee}$ of \cref{sec:two-point} and \cref{app:wti}, whose spinor-loop part is $\Gamma^{(2)ee}|_\tx{spinor}$ of \cref{eq:Gamma-J12}.
After including all quantum effects in the system, however, the full (nonperturbative) two-point function is expected to acquire a nontrivial momentum dependence. 
Its Taylor expansion around vanishing momentum may be written as
\al{
\Paren{\Gamma_{k=0}^{(2)ee}[q;\bar\Phi]}_{\bb\bc}{}^{\nu\rho}
=A_0+A_2q^2+A_4q^4+A_6q^6+\cdots,
}
with loop factors $A_i$.
The vierbein and the LL gauge field mix already through the Einstein--Cartan term, so the full Hessian has the block form
\al{
\Gamma_k^{(2)}[q;\bar\Phi] = \pmat{
\Gamma_{k}^{(2)ee}[q;\bar\Phi] & \Gamma_{k}^{(2)e{\omega}}[q;\bar\Phi]\\[1ex]
\Gamma_{k}^{(2){\omega}e}[q;\bar\Phi] & \Gamma_{k}^{(2){\omega}{\omega}}[q;\bar\Phi]  
}.
}
The propagator poles therefore cannot be read off the low-momentum expansion alone; they require the full (or resummed) Hessian, and after gauge fixing the pole condition is
\al{
\det\Paren{\Gamma_{k=0}^{(2)}[q;\bar\Phi]}=0.
}
Without gauge fixing the determinant would vanish identically on the gauge zero modes, and among the zeros of the gauge-fixed determinant only the gauge-independent ones are physical poles.

In the standard metric formulation, such FRG approaches are known as the form factor approach \cite{Knorr:2019atm,Knorr:2021iwv,Knorr:2022dsx} and the fluctuation approach \cite{Christiansen:2015rva,Pawlowski:2020qer,Bonanno:2021squ,Fehre:2021eob,Pawlowski:2023gym} within the asymptotic safety scenario for quantum gravity.
We will report the explicit computation of the propagator from the FRG elsewhere.
See Refs.~\cite{Daum:2010qt,Harst:2012ni,Dona:2012am,Daum:2013fu,Harst:2014vca,Harst:2015eha} for the asymptotic safety scenario within the first-order formalism.

\section{Mode decomposition of the vierbein two-point function}
\label{app:channels}

The two-point results of \cref{sec:two-point} are index-explicit.
This appendix records their resolution into spin--parity channels, which makes the mode content visible term by term.
The projectors and the mode decomposition of \cref{app:channels-modes} are standard material, collected here in our conventions.
The channel-resolved form of the one-loop divergences in \cref{app:channels-divergences}, and the statements drawn from it, are new.

\subsection{Projectors and the vierbein modes}
\label{app:channels-modes}

On the flat reference~\labelcref{flat reference background}, in momentum space, define the transverse and longitudinal projectors
\al{
P^{\perp}_{\mu\nu}
	&:=	\eta_{\mu\nu}-\frac{q_{\mu}q_{\nu}}{q^{2}},
&
P^{\parallel}_{\mu\nu}
	&:=	\frac{q_{\mu}q_{\nu}}{q^{2}},
	\label{eq:PperpPpar}
}
for $q^{2}:=\eta^{\mu\nu}q_{\mu}q_{\nu}\neq0$.
Their traces are $\eta^{\mu\nu}P^\perp_{\mu\nu}=3$ and $\eta^{\mu\nu}P^\parallel_{\mu\nu}=1$. Since a projector has eigenvalues $0$ and $1$ only, its trace is the dimension of the space it projects onto, so the $4$ components of a vector split into $3$ transverse plus $1$ longitudinal.

On a pair of indices, these assemble into the standard spin projectors of Barnes--Rivers type~\cite{Rivers:1964nfl,Barnes:1965ylk,VanNieuwenhuizen:1973fi}.
The corresponding spin--parity resolution of the torsion field into its $2^{\pm}$, $1^{\pm}$, and $0^{\pm}$ blocks goes back to Ref.~\cite{Hayashi:1980ir}.
The mechanical rule is to distribute $P^{\perp}$ and $P^{\parallel}$ over the two indices, symmetrize or antisymmetrize the pair, and split off the trace wherever one exists:
\al{
P^{(2)}_{\mu\nu,\rho\sigma}
	&:=	\frac12\pn{P^{\perp}_{\mu\rho}P^{\perp}_{\nu\sigma}+P^{\perp}_{\mu\sigma}P^{\perp}_{\nu\rho}}
		-\frac13\,P^{\perp}_{\mu\nu}P^{\perp}_{\rho\sigma},\nn
P^{(1)}_{\mu\nu,\rho\sigma}
	&:=	\frac12\pn{P^{\perp}_{\mu\rho}P^{\parallel}_{\nu\sigma}
		+P^{\perp}_{\mu\sigma}P^{\parallel}_{\nu\rho}
		+P^{\perp}_{\nu\rho}P^{\parallel}_{\mu\sigma}
		+P^{\perp}_{\nu\sigma}P^{\parallel}_{\mu\rho}},\nn
P^{(0)\perp}_{\mu\nu,\rho\sigma}
	&:=	\frac13\,P^{\perp}_{\mu\nu}P^{\perp}_{\rho\sigma},
&
P^{(0)\parallel}_{\mu\nu,\rho\sigma}
	&:=	P^{\parallel}_{\mu\nu}P^{\parallel}_{\rho\sigma},
	\nn
P^{(0)\perp\parallel}_{\mu\nu,\rho\sigma}
	&:=	\frac{1}{\sqrt3}\,P^{\perp}_{\mu\nu}P^{\parallel}_{\rho\sigma},
&
P^{(0)\parallel\perp}_{\mu\nu,\rho\sigma}
	&:=	\frac{1}{\sqrt3}\,P^{\parallel}_{\mu\nu}P^{\perp}_{\rho\sigma},\nn
P^{a\perp}_{\mu\nu,\rho\sigma}
	&:=	\frac12\pn{P^{\perp}_{\mu\rho}P^{\perp}_{\nu\sigma}-P^{\perp}_{\mu\sigma}P^{\perp}_{\nu\rho}},
	\nn
P^{a\parallel}_{\mu\nu,\rho\sigma}
	&:=	\frac12\pn{P^{\perp}_{\mu\rho}P^{\parallel}_{\nu\sigma}
		-P^{\perp}_{\mu\sigma}P^{\parallel}_{\nu\rho}
		-P^{\perp}_{\nu\rho}P^{\parallel}_{\mu\sigma}
		+P^{\perp}_{\nu\sigma}P^{\parallel}_{\mu\rho}},
	\label{eq:BR-projectors}
}
where $P^{(2)}$, $P^{(1)}$, $P^{(0)\perp}$, $P^{(0)\parallel}$ act on the symmetric pair (the metric mode); $P^{a\perp}$ (transverse) and $P^{a\parallel}$ (longitudinal) act on the antisymmetric pair (the linearized NG mode); and $P^{(0)\perp\parallel}$ and $P^{(0)\parallel\perp}$ mix the two scalar traces.
In the traditional lettering in Ref.~\cite{VanNieuwenhuizen:1973fi}, $(P^{(0)\perp},P^{(0)\parallel},P^{a\perp},P^{a\parallel})$ are $(P^{(0s)},P^{(0w)},P^{(b)},P^{(e)})$.\footnote{
In the rest frame, the symmetric pair reads as an energy-momentum tensor and the antisymmetric pair as a field-strength tensor.
The letters $s$ and $w$ stand for the stress and work (energy) scalars, and $b$ and $e$ for the magnetic and electric vectors.
}

The symmetric and antisymmetric fields resolve in these channels term by term,
\al{
\h h_{\mu\nu}
	&=	\pn{P^{(2)}+P^{(0)\perp}+P^{(1)}+P^{(0)\parallel}}_{\mu\nu,\rho\sigma}\,\h h^{\rho\sigma},
	\label{eq:channel-completeness-symmetric}\\
\h a_{\mu\nu}
	&=	\pn{P^{a\perp}+P^{a\parallel}}_{\mu\nu,\rho\sigma}\,\h a^{\rho\sigma},
	\label{eq:channel-completeness-antisymmetric}
}
each pair's projectors summing to the identity on it.\footnote{
The two transfer operators $P^{(0)\perp\parallel}$ and $P^{(0)\parallel\perp}$ are not projectors: Each maps one scalar channel onto the other, so they have no place in the completeness relations.
Their factor $1/\sqrt3$ compensates the traces $3$ of $P^{\perp}$ and $1$ of $P^{\parallel}$, so that the two compose to the scalar projectors, $P^{(0)\perp\parallel}P^{(0)\parallel\perp}=P^{(0)\perp}$ and $P^{(0)\parallel\perp}P^{(0)\perp\parallel}=P^{(0)\parallel}$.
In a two-point function the two scalar channels can mix, and then the transfer operators appear.
They do in \cref{eq:channel-resolved} below, with the coefficient $\sqrt3\,m^{4}$; the $\sqrt3$ cancels the $1/\sqrt3$, and the term is $m^{4}\bigl(P^{\perp}_{\mu\nu}P^{\parallel}_{\rho\sigma}+P^{\parallel}_{\mu\nu}P^{\perp}_{\rho\sigma}\bigr)$.
}
The vierbein fluctuation~\labelcref{eq:ahat} collects the two:
\al{
\h e_{\mu\nu}\fn{x}
	&=	\frac12\,\h h_{\mu\nu}\fn{x}+\h a_{\mu\nu}\fn{x}+\cdots,
	\label{eq:vierbein-split}
}
where we used \cref{eq:hhat}, and the dots denote terms of second and higher order in $\h h$ and $\h a$.
The two projector sets~\labelcref{eq:channel-completeness-symmetric,eq:channel-completeness-antisymmetric} resolve these two parts completely, so together they resolve all sixteen components of a vierbein leg of the 1PI two-point function into six channels.
The remainder of the appendix unpacks the antisymmetric and the symmetric pair in turn, and then resolves the divergences.

Explicitly, the linearized NG mode splits as
\al{
\h a_{\mu\nu}
	&=	P^{\perp}{}_{\mu}{}^{\rho}\,P^{\perp}{}_{\nu}{}^{\sigma}\h a_{\rho\sigma}
		+2\,q_{[\mu}^{}\h\ell_{\nu]}^{},
	\label{eq:NG-split}
}
with
\al{
\h\ell_{\nu}:={q^{\mu}\ov q^{2}}\h a_{\mu\nu}.
}
The first term in \cref{eq:NG-split} is $(P^{a\perp}\h a)_{\mu\nu}$ and the second is $(P^{a\parallel}\h a)_{\mu\nu}$, the resolution~\labelcref{eq:channel-completeness-antisymmetric} made explicit.
The two terms carry $3$ components each. The first is an antisymmetric matrix on the $3$-dimensional transverse space, $3\cdot2\,/2=3$, and the second is carried by $\h\ell_{\nu}$, itself transverse by the antisymmetry of $\h a$ ($q^{\nu}\h\ell_{\nu}=0$), again $3$; together they are the $6$ of the antisymmetric pair.
Projecting both indices transverse and antisymmetrizing gives $P^{a\perp}$ directly, the antisymmetric $3\times3$ tensors being already irreducible ($3$ components).
On the longitudinal part $2\,q_{[\mu}^{}\h\ell_{\nu]}^{}$, the exterior derivative~\labelcref{eq:da} places two factors of the same momentum under one total antisymmetrization and vanishes identically, so the longitudinal channel carries no torsion.

The metric mode resolves likewise,
\al{
\h h_{\mu\nu}
	&=	\pn{P^{\perp}{}_{\mu}{}^{\rho}\,P^{\perp}{}_{\nu}{}^{\sigma}
		-\frac13\,P^{\perp}_{\mu\nu}\,P^{\perp\rho\sigma}}\h h_{\rho\sigma}
		+\frac13\,P^{\perp}_{\mu\nu}\,P^{\perp\rho\sigma}\h h_{\rho\sigma}
		+2\,q_{(\mu}^{}\h c_{\nu)}^{}
		+P^{\parallel}_{\mu\nu}\,\h w,
	\label{eq:metric-split}
}
with
\al{
\h c_{\nu}
	&:=	P^{\perp}{}_{\nu}{}^{\rho}\,\h h_{\rho\sigma}{q^{\sigma}\ov q^{2}},&
\h w
	&:=	{q^{\rho}q^{\sigma}\ov q^{2}}\h h_{\rho\sigma}.
}
The four terms in \cref{eq:metric-split} are, in order, the images of $P^{(2)}$, $P^{(0)\perp}$, $P^{(1)}$, and $P^{(0)\parallel}$, and carry $5+1+3+1$ components, as follows:
The first two terms together form the transverse projection $P^{\perp}{}_{\mu}{}^{\rho}\,P^{\perp}{}_{\nu}{}^{\sigma}\h h_{\rho\sigma}$, which is a symmetric matrix on the $3$-dimensional transverse space, with $3\cdot4\,/2=6$ components.
Of these two, the first is its traceless part, the spin-$2$ mode, with $6-1=5$ components; the second is its trace along $P^{\perp}$, with $1$.
The third term is the transverse vector $\h c_{\nu}$, with $3$ components; the fourth is the scalar $\h w$, with $1$.
Together they are the $10$ of the symmetric pair, and $6+10=16$ completes the count of the polar decomposition~\labelcref{eq:polar-decomposition}.

We explain the transverse-traceless part of \cref{eq:metric-split} in more detail.
On the symmetric pair of indices $\rho\sigma$, the symmetrized transverse projection reads
\al{
\Pi^{\perp}_{\mu\nu,\rho\sigma}
	&:=	\frac12\pn{P^{\perp}_{\mu\rho}P^{\perp}_{\nu\sigma}+P^{\perp}_{\mu\sigma}P^{\perp}_{\nu\rho}}.
	\label{eq:sym-perp}
}
Its full trace is $\eta^{\mu\rho}\eta^{\nu\sigma}\Pi^{\perp}_{\mu\nu,\rho\sigma}=\frac12\pn{3\cdot3+3}=6$, the count of a symmetric $3\times3$ matrix.
This $6$ is reducible. It contains the invariant line spanned by the transverse metric $P^{\perp}_{\mu\nu}$ itself, of norm $P^{\perp}_{\mu\nu}P^{\perp\mu\nu}=P^{\perp\mu}{}_{\mu}=3$; the normalized projector onto that line is the $P^{(0)\perp}$ of \cref{eq:BR-projectors}, its $\frac13$ being nothing but $1/P^{\perp\mu}{}_{\mu}$.
Removing the line from the product leaves the spin-$2$ projector,
\al{
P^{(2)}_{\mu\nu,\rho\sigma}
	&=	\Pi^{\perp}_{\mu\nu,\rho\sigma}-P^{(0)\perp}_{\mu\nu,\rho\sigma}.
	\label{eq:P2-derivation}
}
It is transverse because every index carries a $P^{\perp}$, traceless on each symmetric index pair because $\eta^{\mu\nu}\Pi^{\perp}_{\mu\nu,\rho\sigma}=P^{\perp}_{\rho\sigma}=\eta^{\mu\nu}P^{(0)\perp}_{\mu\nu,\rho\sigma}$, and it carries $6-1=5$ components, the helicities of spin $2$: the transverse-traceless projector of linearized gravity.
One transverse and one longitudinal index give $P^{(1)}$ on the symmetric pair and $P^{a\parallel}$ on the antisymmetric one; two longitudinal indices give $P^{(0)\parallel}$; and $P^{(0)\perp\parallel}$ and $P^{(0)\parallel\perp}$ connect the two traces.

\subsection{Channel-resolved divergences}
\label{app:channels-divergences}

In the channels~\labelcref{eq:BR-projectors}, the divergences~\labelcref{eq:J1(q)-result} and~\labelcref{eq:J2(q)-result} take the form
\al{
-\pn{J_{1}+J_{2}}_{\mu\nu,\rho\sigma}\Big|_{1/\epsilon}
	&=	\frac{1}{16\pi^2\epsilon}\bigg[
		-\pn{m^{4}+\frac13\,m^{2}q^{2}+\frac{1}{10}\,q^{4}}P^{(2)}
		-m^{4}\,P^{(1)}
	\nn
	&\phantom{=	\frac{1}{16\pi^2\epsilon}\bigg[}
		+\pn{2m^{4}+\frac23\,m^{2}q^{2}}P^{(0)\perp}
		+\sqrt3\,m^{4}\pn{P^{(0)\perp\parallel}+P^{(0)\parallel\perp}}
	\nn
	&\phantom{=	\frac{1}{16\pi^2\epsilon}\bigg[}
		+\pn{m^{4}-m^{2}q^{2}-\frac16\,q^{4}}P^{a\perp}
		+m^{4}\,P^{a\parallel}
		\bigg]_{\mu\nu,\rho\sigma},
	\label{eq:channel-resolved}
}
where the frame legs of \cref{eq:J1(q)-result,eq:J2(q)-result} are converted with the flat reference and all legs are lowered with $\eta$.
Each form factor multiplies its own mode of the vierbein resolution~\labelcref{eq:vierbein-split} with \cref{eq:channel-completeness-symmetric,eq:channel-completeness-antisymmetric}.
Several structural statements are immediate:
\begin{itemize}
\item The $P^{(0)\parallel}$ entry vanishes.
On this mode the vierbein fluctuation is $\h e_{\mu\nu}=\frac12\,\h h_{\mu\nu}=\frac12P^{\parallel}_{\mu\nu}\h w\propto q_{\mu}q_{\nu}$, symmetric and purely longitudinal.
Only the cosmological term could reach it, since the quadratic forms of the curvature invariants are transverse and vanish on this mode, while the torsion terms touch only the antisymmetric part.
The cosmological term couples through the vierbein determinant, whose second-order piece is $\frac12\pn{(\h e^{\mu}{}_{\mu})^{2}-\h e^{\mu}{}_{\nu}\h e^{\nu}{}_{\mu}}$.
The trace then brings $q^{\mu}q_{\mu}=q^{2}$, the matrix square brings $q^{\mu}q_{\nu}\,q^{\nu}q_{\rho}=q^{2}\,q^{\mu}q_{\rho}$, so both terms equal $(q^{2})^{2}$ times the same prefactor, and the difference vanishes.
\item No entry connects the symmetric sector with the antisymmetric one.
The only entries connecting two different modes lie within the symmetric sector.
They are the transfer pair between its two scalar traces, $P^{(0)\perp\parallel}$ and $P^{(0)\parallel\perp}$.
These entries are contact terms from the cosmological term.
The determinant term carries no momentum, so it has no reason to respect the split of the scalar sector into its transverse and longitudinal traces, and its scalar block is what links the two.
\item Every momentum-carrying channel ($P^{(1)}$, $P^{(0)\parallel}$, $P^{(0)\perp\parallel}$, $P^{(0)\parallel\perp}$, $P^{a\parallel}$) is a $q$-independent contact term.
This follows from the GC WT identity~\labelcref{eq:wi_wi-3}.
Applied to either external leg, the identity contracts the two-point function with $q_{\nu}$, which annihilates every transverse structure; each scalar transfer operator survives one of the two contractions, so the five channels listed are constrained and no other.
The right-hand side is linear in $q$, with coefficients set by the momentum-independent tadpole~\labelcref{eq:J3(q)-result}.
Together, the two contractions give five independent conditions on the five form factors, which fixes them to the displayed multiples of $m^{4}$.
\item The $m^{2}q^{2}$ terms sit in $P^{(2)}$ and $P^{(0)\perp}$ for the induced Einstein--Hilbert term of \cref{eq:induced-action}, and in $P^{a\perp}$ for the induced torsion mass term.
\item The $q^{4}$ terms sit only in $P^{(2)}$ and $P^{a\perp}$, the channels of the two four-derivative invariants of \cref{eq:induced-action}; the absence of a $q^{4}$ term in $P^{(0)\perp}$ is the vanishing of the divergent $\os g R^{2}$ coefficient.
\item In the massless limit only the two four-derivative terms survive.
For the dimensional-regularization $1/\epsilon$ pole, a massless fermion loop carries no mass scale, so only the two dimensionless couplings of the four-derivative invariants survive.
A physical-cutoff matching can separately contain power-sensitive lower-derivative terms, as discussed in \cref{sec:summary}.
\end{itemize}
Every display of this appendix is verified to machine precision, in this signature, directly from the printed brackets of \cref{eq:J1(q)-result,eq:J2(q)-result}.

\section{Ward--Takahashi identities of the true GC and LL invariance}
\label{app:wti}

In this appendix, we derive the WT identities of the true GC and LL invariance, the GC WT identity and the LL WT identity of the main text, to first order in $\hat{e}$.
They apply to any anomaly-free 1PI effective action, and the invariance of \cref{eq:Suv} and of the Dirac measure under both transformations makes them applicable to the spinor-loop contribution used in the main text.
For that contribution, $\Gamma$ below is $\Gamma_\tx{spinor}$ of \cref{sec:two-point}, and its vertex functions are the loop amplitudes of \cref{eq:Gamma-J12,eq:Gamma-J3}.
We expand the effective action $\Gamma$ at $\hat\omega=0$ in the vierbein fluctuation as
\begin{align}
\label{eq:wti_Gamma}
\Gamma\Big|_{\hat\omega=0}
=
\Gamma^{(0)}
+
(\Gamma^{(1)e})_\mathbf{a}{}^\mu \,
\hat{e}^\mathbf{a}{}_\mu
+
\frac{1}{2}
(\Gamma^{(2)ee})_\mathbf{ab}{}^{\mu\nu}\,
\hat{e}^\mathbf{a}{}_\mu
\hat{e}^\mathbf{b}{}_\nu
+
\mathcal{O}\fn{\hat{e}^3} .
\end{align}
Note that coefficients of the fluctuations, $(\Gamma^{(n)e{\cdots}e})_{\mathbf{a}_1^{}\cdots \mathbf{a}_n^{}}{}^{\mu_1^{}\cdots \mu_n^{}}$,
are symmetrized under the simultaneous exchange of the index pairs $(\mathbf{a}_i^{},\mu_i^{})$; for $n=2$, $(\Gamma^{(2)ee})_\mathbf{ba}{}^{\nu\mu}=(\Gamma^{(2)ee})_\mathbf{ab}{}^{\mu\nu}$.

\subsection{GC WT identity}
\label{app:wti-ld}

The GC invariance is implemented on the fixed chart by the LD transformation of \cref{sec:LD}.
The variation of the fluctuation of the vierbein under the true LD transformation is given by
\begin{align}
\label{eq:wti_delta-ehat}
\delta_\mathrm{LD} \hat{e}^\mathbf{a}{}_\mu
=
-
\Big[
\xi^\nu \p_\nu \bar{e}^\mathbf{a}{}_\mu
+(\p_\mu \xi^\nu)\bar{e}^\mathbf{a}{}_\nu
\Big]
-
\Big[
\xi^\nu \p_\nu \hat{e}^\mathbf{a}{}_\mu
+(\p_\mu \xi^\nu)\hat{e}^\mathbf{a}{}_\nu
\Big] .
\end{align}
The first bracket does not contain $\hat{e}$ and is the inhomogeneous part of the transformation.
The second bracket is linear in $\hat{e}$ and is the homogeneous part.
Taking the variation of the action with respect to $\hat{e}$, the terms linear in $\hat{e}$ become
\begin{align}
\label{eq:wti_dGamma|e}
\delta_\mathrm{LD} \Gamma\Big|_{\mathcal{O}(\hat{e})}
=
\bigg[
-
(\Gamma^{(1)e})_\mathbf{a}{}^\mu \,
\delta_\rho^\nu
+
(\Gamma^{(1)e})_\mathbf{a}{}^\nu \,
\delta_\rho^\mu
+
(\Gamma^{(2)ee})_\mathbf{ab}{}^{\mu\nu}\,
\bar{e}^\mathbf{b}{}_\rho
\bigg]
iq_\nu^{}
\xi^\rho
\hat{e}^\mathbf{a}{}_\mu,
\end{align}
where we assumed the constant background for the vierbein, $\p_\nu \bar{e}^\mathbf{a}{}_\mu = 0$, and where the gauge parameter $\xi$ carries momentum $-q$ and the vierbein fluctuation $\hat{e}$ carries $+q$, the routing used throughout this appendix.\footnote{
As for the $\hat{e}$-independent terms, they involve only the inhomogeneous part of \cref{eq:wti_delta-ehat} and assemble, on the constant background, into the total derivative $-\p_\mu\big[(\Gamma^{(1)e})_\mathbf{a}{}^\mu\,\bar{e}^\mathbf{a}{}_\nu\,\xi^\nu\big]$, which drops for $\xi^\nu$ vanishing at infinity.
}
By requiring that \cref{eq:wti_dGamma|e} vanish for arbitrary $\xi^\rho$, we obtain the GC WT identity to first order in $\hat{e}$,
\begin{align}
\label{eq:wi_wi-Gamma}
q_\nu
(\Gamma^{(2)ee})_\mathbf{ba}{}^{\nu\mu}\,
\bar{e}^\mathbf{b}{}_\rho
=
q_\rho
(\Gamma^{(1)e})_\mathbf{a}{}^\mu \,
-
q_\nu
(\Gamma^{(1)e})_\mathbf{a}{}^\nu \,
\delta_\rho^\mu,
\end{align}
which coincides with \cref{eq:wi_wi-3} at one loop level.

\subsection{LL WT identity}
\label{app:wti-ll}

The true LL transformation acts on the fluctuations with the background held fixed.
With $\Lambda^\ba{}_\bb\fn{x}=\delta^\ba_\bb+\theta^\ba{}_\bb\fn{x}$, the transformation of the vierbein and \cref{eq:LLgaugetrans} give, around the flat background with $\bar{\omega}=0$,
\al{
\label{eq:wti_delta-ll}
\delta_\mathrm{LL}\hat{e}^\mathbf{a}{}_\mu
	&=	\theta^\mathbf{a}{}_\mathbf{b}\,\bar{e}^\mathbf{b}{}_\mu+\theta^\mathbf{a}{}_\mathbf{b}\,\hat{e}^\mathbf{b}{}_\mu, &
\delta_\mathrm{LL}\hat{\omega}^\mathbf{ab}{}_\mu
	&=	-\p_\mu\theta^\mathbf{ab}+\mathcal{O}\fn{\hat{\omega}}.
}
The LL-gauge-field fluctuation enters the LL WT identity through its one-point function and through its mixed two-point function with the vierbein, so we extend the expansion~\labelcref{eq:wti_Gamma} to
\al{
\label{eq:wti_Gamma-ll}
\Gamma
	&=	\Gamma\Big|_{\hat{\omega}=0}
	+\frac{1}{2}(\Gamma^{(1){\omega}})_\mathbf{ab}{}^{\mu}\,\hat{\omega}^\mathbf{ab}{}_\mu
	+\frac{1}{2}(\Gamma^{(2)e{\omega}})_{\mathbf{c},\mathbf{ab}}{}^{\rho\mu}\,\hat{e}^\mathbf{c}{}_\rho\,\hat{\omega}^\mathbf{ab}{}_\mu
	+\mathcal{O}\fn{\hat{\omega}^2,\hat{e}^2\hat{\omega}},
}
where the first term is the expansion~\labelcref{eq:wti_Gamma}.

Under \cref{eq:wti_delta-ll} at $\hat{\omega}=0$, the terms of $\delta_\mathrm{LL}\Gamma$ linear in $\hat{e}$ are\footnote{
The $\hat{e}$-independent terms are $(\Gamma^{(1)e})_\mathbf{a}{}^\mu\,\theta^\mathbf{a}{}_\mathbf{b}\,\bar{e}^\mathbf{b}{}_\mu$, which vanishes because background Lorentz covariance makes $(\Gamma^{(1)e})_\mathbf{a}{}^\mu\propto\bar{e}_\mathbf{a}{}^\mu$ on the flat background, as \cref{eq:J3(q)-result} exhibits, while $\theta$ is antisymmetric, and $-\frac{1}{2}(\Gamma^{(1){\omega}})_\mathbf{ab}{}^{\mu}\,\p_\mu\theta^\mathbf{ab}$, a total derivative on the constant background.
}
\al{
\label{eq:wti_dGamma-ll}
\delta_\mathrm{LL}\Gamma\Big|_{\mathcal{O}(\hat{e})}
	&=	\bigg[
	(\Gamma^{(1)e})_\mathbf{a}{}^\rho\,\eta_\mathbf{bc}
	+(\Gamma^{(2)ee})_\mathbf{ac}{}^{\nu\rho}\,\bar{e}_{\mathbf{b}\nu}
	+\frac{i}{2}q_\nu\,(\Gamma^{(2)e{\omega}})_{\mathbf{c},\mathbf{ab}}{}^{\rho\nu}
	\bigg]\theta^\mathbf{ab}\hat{e}^\mathbf{c}{}_\rho .
}
Only the part of the bracket antisymmetric in $\mathbf{ab}$ multiplies $\theta^\mathbf{ab}$, and the antisymmetrization doubles the mixed term, whose coefficient is already antisymmetric.
Requiring \cref{eq:wti_dGamma-ll} to vanish for arbitrary $\theta^\mathbf{ab}$, we obtain the LL WT identity to first order in $\hat{e}$,
\al{
\label{eq:wi_ll-Gamma}
(\Gamma^{(2)ee})_\mathbf{ac}{}^{\nu\rho}\,\bar{e}_{\mathbf{b}\nu}
-(\Gamma^{(2)ee})_\mathbf{bc}{}^{\nu\rho}\,\bar{e}_{\mathbf{a}\nu}
+iq_\nu\,(\Gamma^{(2)e{\omega}})_{\mathbf{c},\mathbf{ab}}{}^{\rho\nu}
	&=	\eta_\mathbf{ac}\,(\Gamma^{(1)e})_\mathbf{b}{}^\rho
	-\eta_\mathbf{bc}\,(\Gamma^{(1)e})_\mathbf{a}{}^\rho .
}
The mixed term is absent at $q=0$:
No Lorentz-invariant constant tensor carries the five indices of $(\Gamma^{(2)e{\omega}})_{\mathbf{c},\mathbf{ab}}{}^{\rho\nu}$, so the mixed two-point function is at least linear in $q$.
The mixed term of \cref{eq:wi_ll-Gamma} is this function times $q_\nu$, which comes from the derivative $\p_\nu\theta^\mathbf{ab}$ in \cref{eq:wti_delta-ll}; it is therefore at least quadratic in $q$.
At $q=0$ the identity thus relates the $ee$ two-point function to the tadpole alone.
At one loop, \cref{eq:wi_ll-Gamma} becomes \cref{eq:wi_ll-3}, where $\Gamma^{(2)ee}=J_1+J_2$ and $\Gamma^{(1)e}=J_3$ by \cref{eq:Gamma-J12,eq:Gamma-J3}, and $\Gamma^{(2)e{\omega}}$ is the mixed amplitude $J^{e\omega}$ of \cref{eq:wi_ll-3}.

\section{Polar decomposition of the vierbein}
\label{app:polar}

This appendix establishes the polar factorization~\labelcref{eq:polar-decomposition}: the uniqueness of the decomposition, the expansions of the fluctuation modes, and the transformation laws that underlie the all-orders removal of the would-be NG boson stated in \cref{sec:trade}.

Uniqueness near the background follows from the linearization.
For the compound matrix $e^{\ba}{}_{\bb}\fn{x}:=e^{\ba}{}_{\mu}\fn{x}\,\bar e_{\bb}{}^{\mu}\fn{x}$, the factorization~\labelcref{eq:polar-decomposition} is the map $\pn{\h\kappa,\sth}\mapsto e^{\h\kappa}\pn{\delta+\frac12\sth}$, which linearizes at the background to $\delta+\h\kappa+\frac12\sth$.
A matrix splits uniquely into its antisymmetric and symmetric parts, so the linearization is invertible, and the map is a local diffeomorphism onto a neighborhood of the background.
Every vierbein near the background therefore factors as in \cref{eq:polar-decomposition}, with unique $\pn{\h\kappa,\sth}$.

Expanded in the polar fields, the fluctuation modes in \cref{eq:ahat} read
\al{
\h a_{\mu\nu}\fn{x}
	&=	\h\kappa_{\mu\nu}\fn{x}
		+\frac14\Pn{\h\kappa_{\mu}{}^{\rho}\fn{x}\,\sth_{\rho\nu}\fn{x}
			+\sth_{\mu}{}^{\rho}\fn{x}\,\h\kappa_{\rho\nu}\fn{x}}
		+\dotsb,
	\label{eq:polar-linear}\\
\h s_{\mu\nu}\fn{x}
	&{}=	\frac12\,\sth_{\mu\nu}\fn{x}
		+\frac14\Pn{\h\kappa_{\mu}{}^{\rho}\fn{x}\,\sth_{\rho\nu}\fn{x}
			-\sth_{\mu}{}^{\rho}\fn{x}\,\h\kappa_{\rho\nu}\fn{x}}
		+\frac12\,\h\kappa_{\mu}{}^{\rho}\fn{x}\,\h\kappa_{\rho\nu}\fn{x}
		+\dotsb,
	\label{eq:polar-symmetric}
}
where the ellipses collect the terms of third and higher order in $(\h\kappa,\sth)$, and both series continue at every order.
The metric combination~\labelcref{eq:hhat} nevertheless collapses.
The Lorentz element preserves $\eta$ and cancels in the composite metric~\labelcref{composite metric},
\al{
g_{\mu\nu}\fn{x}
	&=	\pn{\delta^{\rho}_{\mu}+\frac12\sth{}^{\rho}_{\mu}\fn{x}}
		\bar g_{\rho\sigma}\fn{x}
		\pn{\delta^{\sigma}_{\nu}+\frac12\sth{}^{\sigma}_{\nu}\fn{x}},
	\label{eq:polar-metric}
}
which is the multiplicative form of the exact relation~\labelcref{eq:stretch-hhat}, so the metric fluctuation terminates at the second order in the stretch and is free of the would-be NG boson at every order.

Two independent rotations act on the decomposition: the LL transformation $e^{\ba}{}_{\mu}\fn{x}\to\Lambda^{\ba}{}_{\bb}\fn{x}\,e^{\bb}{}_{\mu}\fn{x}$ of the vierbein at fixed background, and a relabeling of the background frame, $\bar e^{\,\ba}{}_{\mu}\fn{x}\to\bar\Lambda^{\ba}{}_{\bb}\fn{x}\,\bar e^{\,\bb}{}_{\mu}\fn{x}$.
Uniqueness of the decomposition~\labelcref{eq:polar-decomposition} places both on the Lorentz factor,
\al{
\pa{e^{\h\kappa\fn{x}}}^{\ba}{}_{\bb}
	&\to	\Lambda^{\ba}{}_{\bc}\fn{x}\,\pa{e^{\h\kappa\fn{x}}}^{\bc}{}_{\bd}\,\pa{\bar\Lambda^{-1}\fn{x}}^{\bd}{}_{\bb}.
	\label{eq:polar-bifundamental}
}
The Lorentz factor is a bifundamental link element connecting the dynamical frame with the background frame, and its exponent $\h\kappa$ is the would-be NG boson to all orders; the antisymmetric fluctuation~$\h a$ coincides with it at linear order~\labelcref{eq:polar-linear}.
The stretch $\sth_{\mu\nu}$ carries no frame leg at all. It is exactly inert under both rotations, an exact LL singlet, transforming only as a GC tensor.

Because the LL gauge transformation acts on the Lorentz factor alone, it can be used to set $\h\kappa\fn{x}=0$ at every order.
In this gauge the fluctuation is the stretch alone, so $\h a$ vanishes exactly, while the metric fluctuation~\labelcref{eq:stretch-hhat}, which depends only on the stretch, is unchanged.
This establishes the unitary gauge of \cref{sec:trade}, in which the LL orientation is absorbed into the St\"uckelberg combination with the connection.

\section{Heat-kernel cross-check of the induced action}
\label{app:heatkernel}

This appendix reproduces the five coefficients of \cref{eq:induced-action} from the Seeley--DeWitt expansion of the squared Dirac operator, without the diagrams of \cref{sec:two-point} or any gauge fixing.
We work in Euclidean signature, $\anticommutator{\gamma^{\mu}}{\gamma^{\nu}}=2\delta^{\mu\nu}\I$, and compare the $1/\epsilon$ pole.
For the mostly-plus metric~\labelcref{eq:tangentspacemetric} the Wick rotation is a change of coordinates, so the relative coefficients in \cref{eq:induced-action} are signature independent.
The overall sign is not, because the Euclidean action density is minus the Lorentzian Lagrangian density.
The Euclidean pole~\labelcref{eq:hk-map} below is therefore minus \cref{eq:induced-action}, and the two brackets are compared term by term in \cref{tab:hk}.
The first row is the one-loop vacuum energy of a Dirac fermion, $+m^{4}/16\pi^{2}\epsilon$, which enters \cref{eq:induced-action} as the shift $\Lambda_\tx{cc}\to\Lambda_\tx{cc}+m^{4}/16\pi^{2}\epsilon$ of the bare term in \cref{eq:Suv}.

Integrating out the spinor on the background $(\bar e,\bar\omega)$ gives the Euclidean effective action $\Gamma^\tx{E}_\tx{spinor}=-\ln\det\pn{\slashed{\mc{D}}+m}=-\tfrac12\ln\det\pn{\Delta+m^{2}}$, where $\slashed{\mc D}=\gamma^{\mu}\mc D_{\mu}$ is the Dirac operator of \cref{eq:Suv} and
\al{
\Delta	&:=	-\slashed{\mc{D}}^{2}=-\pn{\os\varpi\nabla^{\mu}\os\varpi\nabla_{\mu}+E},	&
\os\varpi\nabla_{\mu}	&:=	\os g\nabla_{\mu}+\varpi_{\mu},	&
\Omega_{\mu\nu}	&:=	\commutator{\os\varpi\nabla_{\mu}}{\os\varpi\nabla_{\nu}},
	\label{eq:hk-laplace}
}
with the connection $\varpi_{\mu}$ and the matrix-valued function $E$ read off from $\slashed{\mc{D}}^{2}$.
In $D=4-2\epsilon$ the proper-time representation with $\operatorname{Tr}e^{-t\Delta}=\pn{4\pi t}^{-D/2}\sum_{k\ge0}t^{k}\int\abb{e}\,b_{2k}$ gives the pole
\al{
\Gamma^\tx{E}_\tx{spinor}\Big|_{1/\epsilon}
	&=	{1\ov16\pi^{2}\epsilon}\int\df^{4}x\,\abb{e}
		\pn{{m^{4}\ov4}\,b_{0}-{m^{2}\ov2}\,b_{2}+{1\ov2}\,b_{4}},
	\label{eq:hk-map}
}
with the Seeley--DeWitt coefficients~\cite{Vassilevich:2003xt} $b_{0}=\operatorname{tr}\I$, $b_{2}=\operatorname{tr}\pn{E+\tfrac16\os gR\,\I}$, and
\al{
b_{4}	&=	\frac1{360}\operatorname{tr}\Bigl[60\,\os gR\,E+180\,E^{2}+30\,\Omega_{\mu\nu}\Omega^{\mu\nu}
		+\pn{5\,\os gR^{2}-2\,\os gR_{\mu\nu}\os gR^{\mu\nu}+2\,\os gR_{\mu\nu\rho\sigma}\os gR^{\mu\nu\rho\sigma}}\I\Bigr],
	\label{eq:hk-gilkey}
}
where the trace runs over spinor indices and total derivatives are dropped.

On a torsion-free background, the Lichnerowicz identity gives $E=-\tfrac14\os gR\,\I$, and $\Omega_{\mu\nu}=\tfrac12\os gR_{\mu\nu\ba\bb}\sigma^{\ba\bb}$.
The coefficients $b_{2k}$ are then
\al{
b_{0}	&=	4,	&
b_{2}	&=	-\frac13\os gR,	&
b_{4}	&=	\frac1{360}\pn{-18\,\os gC^{2}+11\,E_\tx{GB}},
	\label{eq:hk-metric}
}
where $E_\tx{GB}:=\os gR_{\mu\nu\rho\sigma}\os gR^{\mu\nu\rho\sigma}-4\os gR_{\mu\nu}\os gR^{\mu\nu}+\os gR^{2}$ is the Gauss--Bonnet density.

\begin{table}[t]
\centering
\begin{tabular}{@{}llll@{}}
\toprule
Term in \cref{eq:induced-action} & Coefficient & Heat-kernel input & Literature \\
\midrule
$m^{4}$ & $1$ & $\tfrac14 b_{0}$, $b_{0}=4$ & \cite{Sakharov:1967pk} \\
$m^{2}\os gR$ & $\tfrac16$ & $-\tfrac12 b_{2}$, $b_{2}=-\tfrac13\os gR$ & \cite{Sakharov:1967pk} \\
$\os gC^{2}$ & $-\tfrac1{40}$ & $\tfrac12 b_{4}$, $b_{4}\supset-\tfrac1{20}\os gC^{2}$ & \cite{Birrell:1982ix} \\
$m^{2}\bar K_{[\lambda\mu\nu]}\bar K^{[\lambda\mu\nu]}$ & $-\tfrac32$ & $-\tfrac12 b_{2}$, $b_{2}\supset3\bar K^{2}$ & \cite{Obukhov:1982da,Buchbinder:1985ux,deBerredo-Peixoto:1999qfk,Shapiro:2001rz} \\
$\pn{\os{\bar g}\nabla\bar K_{[\lambda\mu\nu]}}^{2}$ & $-\tfrac14$ & $\tfrac12 b_{4}$, $b_{4}\supset-\tfrac12\pn{\nabla\bar K}^{2}$ & \cite{Obukhov:1982da,Buchbinder:1985ux,deBerredo-Peixoto:1999qfk,Shapiro:2001rz} \\
\bottomrule
\end{tabular}
\caption{The five coefficients in the bracket of \cref{eq:induced-action}, equal to those in the bracket of the Euclidean pole~\labelcref{eq:hk-map}, against the Seeley--DeWitt coefficients of the squared Dirac operator. $\bar K^{2}$ and $(\nabla\bar K)^{2}$ abbreviate the totally antisymmetric contractions displayed in \cref{eq:hk-torsion}.}
\label{tab:hk}
\end{table}

On a flat vierbein background $\bar e^\ba{}_\mu\fn{x}=\delta^\ba_\mu$ with a background LL gauge field $\bar\omega_{\ba\bb\mu}\fn{x}=\bar K_{\ba\bb\mu}\fn{x}$, only the totally antisymmetric part $\bar K_{[\lambda\mu\nu]}$ enters the spinor loop, as shown in \cref{sec:result}.
The dual $S^{\mu}:=\frac1{3!}\ep{\mu\nu\rho\sigma}\bar K_{[\nu\rho\sigma]}$ of $\bar K_{[\lambda\mu\nu]}$ satisfies $S_{\mu}S^{\mu}=\frac16\,\bar K_{[\lambda\mu\nu]}\bar K^{[\lambda\mu\nu]}$, and the coupling is axial,
\al{
\gamma^{\mu}\,\frac12\bar K_{[\ba\bb\mu]}\sigma^{\ba\bb}
	&=	c_{A}\,\gamma_{5}\slashed{S},	&
c_{A}^{2}	&=	\frac94,
	\label{eq:hk-axial}
}
where the sign of $c_{A}$ drops out of the traces below.
Squaring $\slashed{\mc{D}}=\gamma^{\mu}\p_{\mu}+c_{A}\gamma_{5}\slashed{S}$ gives $\varpi_{\mu}=-2c_{A}\gamma_{5}\sigma_{\mu\nu}S^{\nu}$ and
\al{
E	&=	-c_{A}\gamma_{5}\gamma^{\mu}\gamma^{\nu}\p_{\mu}S_{\nu}-c_{A}^{2}S_{\mu}S^{\mu}\,\I-\p^{\mu}\varpi_{\mu}-\varpi^{\mu}\varpi_{\mu},
	\label{eq:hk-E}
}
as in the Riemann--Cartan expansions of Refs.~\cite{Kimura:1981zr,Yajima:2000fu}.
The trace of $E$ gives $b_{2}=\operatorname{tr}E=8c_{A}^{2}S_{\mu}S^{\mu}=3\,\bar K_{[\lambda\mu\nu]}\bar K^{[\lambda\mu\nu]}$.
At quadratic order in $S$, the derivative term of $b_{4}$ comes from $\operatorname{tr}\Omega_{\mu\nu}\Omega^{\mu\nu}$ alone,
\al{
b_{4}\supset\frac1{12}\operatorname{tr}\Omega_{\mu\nu}\Omega^{\mu\nu}
	=	-\frac43c_{A}^{2}\,\p_{\rho}S_{\mu}\p^{\rho}S^{\mu}
	=	-\frac12\,\os{\bar g}\nabla_{\rho}\bar K_{[\lambda\mu\nu]}\os{\bar g}\nabla^{\rho}\bar K^{[\lambda\mu\nu]},
	\label{eq:hk-torsion}
}
because every derivative term of $\operatorname{tr}E^{2}$ carries $\p_{\mu}S^{\mu}$, which vanishes for the exact form~\labelcref{eq:da}.
The two torsion terms, $b_{2}$ and \cref{eq:hk-torsion}, agree with the one-loop axial-torsion divergences of a Dirac fermion in Refs.~\cite{Obukhov:1982da,Obukhov:1983mm,Buchbinder:1985ux,deBerredo-Peixoto:1999qfk}, collected in Eq.~(3.15) of Ref.~\cite{Shapiro:2001rz}.
Their conventions differ from ours by a factor of two in the pole prefactor, a sign per power of curvature, and the Wick sign of the massless derivative term; with these three factors the coefficients agree sector by sector.
Their unit-weight torsion is twice ours, so $T_{[\lambda\mu\nu]}=-2\bar K_{[\lambda\mu\nu]}$ by \cref{torsion fluctuation} and $\ep{\mu\nu\rho\sigma}T_{\nu\rho\sigma}=-12S^{\mu}$.
Their axial vector is this combination with a sign set by their orientation and signature conventions, and $\eta^{2}=c_{A}^{2}/144=\tfrac1{64}$ for their minimal coupling $\eta$.

Inserted in \cref{eq:hk-map}, the coefficients~\labelcref{eq:hk-metric,eq:hk-torsion} give the five terms of \cref{eq:induced-action} with the coefficients listed in \cref{tab:hk}.
They give no $\os gR^{2}$ term, and the Gauss--Bonnet density they contain is invisible to the two-point function.
The agreement is an independent check of the logarithmic coefficients extracted in \cref{sec:two-point}.

\bibliographystyle{JHEP}
\bibliography{refs}
\end{document}